\documentclass[11pt,twoside,onecolumn]{article}
\usepackage[dvips]{epsfig}
\newcommand{\be}{\begin{eqnarray}}
\newcommand{\ee}{\end{eqnarray}}
\newcommand{\bra}[1]{\mbox{$\langle\, #1 \mid$}}

\newcommand{\ket}[1]{\mbox{$\mid #1\,\rangle$}}

\newcommand{\pro}[2]{\mbox{$\langle\, #1 \mid #2\,\rangle$}}

\renewcommand{\natural}{\mbox{{\rm I\hspace{-2truemm} N}}}

\renewcommand{\a}{\hat a}
\newcommand{\ac}{\hat a^{\dagger}}
\renewcommand{\b}{\hat b}
\newcommand{\bc}{\hat b^\dagger}
\title{Gravitational Collapse of a Radiating Shell}
\author{G.L. Alberghi\thanks{e-mail: alberghi@bo.infn.it}$\ ^{,a}$,\
R. Casadio\thanks{e-mail: casadio@bo.infn.it}$\ ^{,a}$,\
G.P. Vacca\thanks{e-mail: vacca@bo.infn.it}$\ ^{,a,b}$
\ and G. Venturi\thanks{e-mail: armitage@bo.infn.it}$\ ^{,a}$\\
 \\
{\em $^a\,$Dipartimento di Fisica, Universit\`a di Bologna and} \\
{\em Istituto Nazionale di Fisica Nucleare,
Sezione di Bologna,}\\
{\em via Irnerio 46, 40126 Bologna, Italy}\\
\\
{\em $^b\,$II Institut f\"ur Theoretische Physik,
Universit\"at Hamburg,}\\
{\em Luruper Chaussee 149, D-22761 Hamburg}}
\begin{document}
\maketitle
\begin{abstract}
We study the collapse of a self-gravitating and radiating shell of
bosonic matter.
The matter constituting the shell is quantized and the construction
is viewed as a semiclassical model of possible black hole formation.
It is shown that the shell internal degrees of freedom are excited
by the quantum non-adiabaticity of the collapse and, consequently,
on coupling them to a massless scalar field, the collapsing matter
emits a burst of coherent (thermal) radiation.
The backreaction on the trajectory is also estimated.
\end{abstract}
\pagestyle{plain}
\raggedbottom
\setcounter{page}{1}
\section{Introduction}
\setcounter{equation}{0}
Much effort has been dedicated to the study of the classical
dynamics of gravitationally collapsing bodies and a fairly
comprehensive understanding of the main features of this
phenomenon is now available (see, {\em e.g.}, \cite{review}
and References therein).
However, it is also clear that classical physics is not
sufficient for a complete description, and for diverse
reasons:
first of all, the predicted point-like singularity which
emerges as the final fate of the collapse is quantum
mechanically unacceptable, just on the basis of naive
consideration of the uncertainty principle; secondly, as
soon as the collapsing body approaches its own gravitational
radius, Hawking radiation \cite{hawking} switches on
and its backreaction should be included properly \cite{haji}.
Moreover, the above two issues are connected, since Hawking's
effect violates the positive energy condition which is a
basic hypothesis of the singularity theorems \cite{pos}.
One therefore expects corrections to the classical picture
already at the semiclassical level, {\em i.e.}, in the region
where matter is properly evolved by quantum equations on a
a space-time whose dynamics is still reliably approximated
by classical equations.
\par
The semiclassical limit for various models has been previously
investigated \cite{cv,qg96,cfv,fvv,shell} by employing a
Born-Oppenheimer (BO) decomposition of the corresponding
minisuperspace wavefunction \cite{bv,bfv,ebo}.
In particular, in Ref.~\cite{shell} the case of a (thin)
shell of quantized scalar matter collapsing in vacuum was
investigated and it was shown that the collapse induces
the production of matter quanta.
Further, explicit conditions were found beyond which the
semiclassical approximation breaks down.
Such a model is, however, both unrealistically simple and
of little physical use, since the absence of any signal from
the shell precludes an observer from witnessing the process.
In order to overcome the latter shortcoming, an effective
minisuperspace action for radiating shells was derived in
Ref.~\cite{vaidya} as the first step in the modelling of
a more realistic case.
This was expected to be useful both for the conceptual
problem of unambiguously quantizing the system and for
obtaining more predictive conclusions:
in fact, the possibility of adding an observer ({\em e.g.},
in the form of a detector coupled to the outgoing radiation)
allows one to define physical (and not just formal)
observables \cite{qg99}.
\par
As mentioned above, one of the most intriguing aspects
of collapsed bodies is the outgoing flux of thermal
radiation predicted by Hawking \cite{hawking}.
This effect is usually studied in the background of a
preexisting black hole, thus separating the problem of
the collapse from the understanding of the onset of the
thermal radiation.
Such an approach is inspired by Hawking's original
computation, where any backreaction on the chosen
Schwarzschild background is neglected and the matter
collapsed to form the black hole plays no role, and is
further supported by the smallness of the (renormalized)
energy-momentum tensor of the radiation in the vicinity
of the horizon \cite{birrell}.
In this framework, one can think of a particle-antiparticle pair
being generated outside the (event \cite{haw} or apparent
\cite{haji}) horizon, with the positive energy particle escaping
in the form of thermal radiation and the negative energy
antiparticle falling inside the horizon decreasing the ADM mass
({\em i.e.}, proper mass plus gravitational energy) of the
singularity.
It is therefore the horizon, a purely geometrical concept,
which appears as the key ingredient for this process, and much
attention is devoted to studying event horizons as if possessing
physical degrees of freedom of their own \cite{bek,membrane}
(see also Refs.~\cite{carlip,ashtekar} for more recent
developments).
However, on recalling that the central singularity is a
transmutation of the collapsed body, one is alternatively
tempted to define the Hawking effect as a transfer of energy
from the ADM mass of the collapsing matter to the radiation
field.
\par
The latter statement can be put on a firmer ground if one
considers the collapse and the onset of the Hawking radiation in
the same and only picture, which is also the scheme we are
naturally led to use if we take the point of view of a (distant)
static observer.
For such an observer the collapse would never end (classically)
and the final fate of the collapsing body necessarily overlaps
with the onset of thermal emission.
Hence, the static observer does not see an intermediate stage
at which there is a true black hole, because the infalling flux
of negative energy annihilates against the (still) collapsing
matter, and it is clear that he would consider the Hawking
radiation as energy lost by the collapsing body when it approaches
its own gravitational radius.
In this scenario, the applicability of the no-hair theorems becomes
questionable, with geometry playing
(at most) a subsidiary role, and one should be able to describe
the whole process in terms of dynamical quantities, in much the
same fashion as is the case for non-gravitational interactions
\cite{thooft}, and possibly recover unitarity \cite{hajicek,h2} .
This approach was explored long ago in Refs.~\cite{gerlach},
whose authors were led to deny the very formation of
black holes (horizons), and has been revived recently from purely
kinematical arguments in the framework of optical geometry
\cite{sonego}.
To summarize the present situation, it still sounds fair to say,
as in Ref.~\cite{haji}, that the issue of whether the horizon
forms or not requires a better understanding of the (quantum
mechanics of the) gravitational collapse.
\par
Instead of constructing a general formalism, in the present paper
we shall continue our study of a specific (quantum mechanical) model
of collapsing bosonic matter, to wit, the {\em self-gravitating\/}
shell (for a purely classical treatment see, {\em e.g.},
Ref.~\cite{tomimatsu}).
Given its wide flexibility as a building block for more complex
configurations, we believe that the conclusions we are able to draw
are sufficiently general to be taken as hints for other situations.
One may query our use of a scalar field, rather than fermionic matter,
for the collapse.
Clearly, in our case one will have, besides some simplifications,
effects associated with the matter's Bose-Einstein nature, for
example the formation of a condensate.
Nonetheless, even if apparently unrealistic, considerable effort
has been dedicated to the effects of such matter ({\em e.g.}, in
the context of boson stars~\cite{mielke}).
\par
Returning to our previous study of a collapsing self gravitating
shell, we consider a ``macro\-shell'' constructed from a large number
of ``microshells'', each of which corresponds to the $s$-wave collapse
of a scalar particle of typical hadronic mass ($m_h\sim 1\,$GeV).
In Ref.~\cite{shell}, we discussed how such microshells were bound
together to form the macroshell and found the confining potential to
be linear and dependent on the shell radius.
This implied that as the macroshell collapses, the confined microshells
undergo non-adiabatic transitions from the ground to higher excited
states.
The attitude we adopted is that, because of their bosonic nature, the
microshells formed a ``condensate'' and that, once enough microshells
were excited (thus widening the macroshell) they would collectively
decay to the ground state by creating additional particles
(in analogy with hadronic string theory wherein a long string -- excited
hadron decays into short strings -- lighter hadrons).
Thus during the collapse the creation of additional microshells
(particles) leads to a backreaction slowing the macroshell.
\par
It is clear that in the above the Schwarzschild (ADM) mass of the shell
does not change.
We previously indicated that a more realistic approach would be to
consider non-adiabatic effects leading to transitions from the ground
to the higher (excited) states for some of the bound microshells and,
subsequently, these states would decay and the microshells would return
to the ground state by emitting radiation \cite{vaidya,qg99}.
At the end of such a process, the proper energy of the macroshell would
be essentially unchanged (except for the small change in ground state
energy due to the change in macroshell radius) and the net balance
is then a transfer of the gravitational energy of the macroshell to the
radiation field which decreases the ADM mass of the system and modifies
the trajectory for the radius of the macroshell.
\par
In our present approach we do not consider particle (microshell)
creation, which would require much greater energy than the excitations
of the microshells due to non-adiabatic effects (in Ref.~\cite{shell}
we also showed that microshell creation occurs for rather extreme cases,
in which the size of the horizon is comparable to the Compton
wavelength of the microshells), but study the latter excitations.
In order to determine the latter excited levels we must describe the
motion of one microshell in the mean gravitational field of the others
and obtain an effective Schr\"odinger equation for the microshell.
Such a study is attempted in the next Section.
Subsequently one must determine the amplitude for the excitation of
a microshell, as a consequence of the time variation of the binding
potential due to the collapse of the shell (this is attempted in
Section~\ref{s_fqp}).
The model is constructed bearing in mind a number of
constraints:
firstly, since our scalar particles have a Compton wavelength of
$\sim 10^{-14}\,$cm, one must have both that the radius of the shell
and the thickness of the macroshell be much greater, otherwise
fluctuations due to the quantum nature of matter will destroy our
semiclassical description of the macroshell motion
(see the consistency conditions for the semiclassical approximation
in our previous paper, Ref.~\cite{shell}).
This essentially implies that the number of microshells of hadronic
mass be at least $10^{40}$ so that, for example, the gravitational
radius is sufficiently large.
Another constraint which will allow us to simplify our results
is to suppose that the collapse be non-relativistic ({\em i.e.},
``slow'' all the way down to the horizon), which implies that only
the first microshell excited states will be relevant.
\par
In the remainder of Section~\ref{f&t} we address the emission of
the radiation (scalar massless particles -- we shall subsequently
just call them photons) and the macroshell trajectory with
backreaction.
Again we shall enforce some constraints, in particular that the
coupling of the radiation to the microshells be sufficiently large
so that the decay time is smaller than the unperturbed collapse time.
Actually, we shall see that such is already the case for a very
small coupling constant.
Further, interesting features of the model are that the collapse
is slow and the shell tickness remains much smaller than the typical
wavelength of the emitted quanta, so that many emissions occur
practically in phase, thus rendering the radiation process highly
coherent and the backreaction great.
As a consequence, it will be sufficient for us to just consider first
order perturbation theory in the radiation coupling constant
and keep terms to lowest order in the macroshell velocity.
\par
One may wonder if our model is useful in describing Hawking's
radiation, which has two main features: the first one is the
thermal spectrum and the second one is its ``observer dependence''.
The former follows from the adiabatic hypothesis that the ADM
mass of the source (black hole) changes slowly in time and,
therefore, one does not necessarily expect to recover a thermal
spectrum in a fully dynamical context in which the backreaction is
included.
By the latter we mean that a freely falling observer remains in
its initial state all the time and should thus not experience
the flux of outgoing energy measured by a static observer, regardless
of the adiabatic approximation.
We just point out that our model shows such a sort of
``observer dependence'', since the proper mass, which remains
constant, is the energy as measured by the observer comoving
with the shell which, in turn, does not experience any change.
At the same time there is a flux of outgoing radiation associated
with the decrease of the ADM (Bondi) mass.
However, whether such a radiation can be consistently identified
with Hawking's or should be considered as a totally different effect
will need further inspection.
Let us note in any case that, in contrast with Hawking radiation,
the coupling constant of the radiation to the microshells will
appear in the final expressions.
\par
Naturally our results will also be constrained by our availability
of computing power: bearing all the above points (and approximations)
in mind, our results are illustrated and summarized in the Conclusions.
\par
We use units for which $c=1$ but explicitly show both Newton's
constant $G$ and Planck's constant $\hbar$.
The Planck length and mass are then given by $\ell_p^2=\hbar\,G$
and $m_p=\hbar/\ell_p$.
\section{The macroshell inner structure}
\setcounter{equation}{0}
\label{structure}
In order to study realistically the evolution of a collapsing
body it is important not to neglect its spatial structure.
For instance, the equivalence principle implies that a point-like
freely falling observer does not experience any gravitational
effect.
However, any object has a spatial extension and tidal forces
become effective when this size gets close to the typical scale
of variation of the space-time curvature.
At the same time, it is also important to keep the model simple
so as to allow one to study it.
A good compromise between realism and simplicity is given by
the {\em macroshell\/} which was introduced in Ref.~\cite{shell}
and which we now review and further develop.
\par
We view the collapsing shell not just as a singular spherical
surface in space \cite{israel} but as a set of a large number
$N$ of {\em microshells\/} which correspond to collapsing
$s$-wave bosons and are described by a wavefunction (thus the
microshell radius is actually the ``average'' radius).
Each microshell has negligible thickness and proper mass $m$
such that their total proper mass equals the proper mass of
the macroshell, $N\,m=M$.
If we denote by $r\in(0,+\infty)$ the usual ``areal'' radial
coordinate, we can order the microshells according to their
area, so that $r_1<r_2<\ldots<r_N$, and assume that the thickness
$\delta$ of the macroshell is small,
\be
0<\delta\equiv r_N-r_1\ll r_1
\ .
\ee
The space between each two microshells has Schwarzschild geometry,
\be
ds^2_i=-\left(1-{2\,G\,M_i\over r}\right)\,dt_i^2
+\left(1-{2\,G\,M_i\over r}\right)^{-1}\,dr^2
+r^2\,d\Omega^2
\ ,
\label{metric}
\ee
with a suitable ADM mass $M_i<M_{i+1}$, $M_0=0$ for $r<r_1$ and
$M_N\equiv M_s$ (total ADM mass) for $r>r_N$.
For this description to be consistent, one must check that the
microshells do not spread during the collapse and
$\delta$ remains small in the sense specified above.
This was already shown in Ref.~\cite{shell} to be ensured by the
mutual gravitational interaction of the microshells if $N$
is sufficiently large and the average radius of the shell $R$ is
much larger then the gravitational radius $R_H\equiv 2\,G\,M_s$.
We shall provide further evidence for this in the following.
\par
Before studying the dynamics of the microshells, one
must also ``fix the gauge'' corresponding to the freedom of making
coordinate transformations on the relevant solution (\ref{metric})
of the Einstein field equations.
This can be done by choosing an ADM foliation \cite{mtw} for the
corresponding space-time manifold.
Two options which cover the domain of outer communication (the region
outside the horizon)
are given by the hypersurfaces $\{\Sigma_\tau\}$
of constant proper time $\tau$, associated with observers comoving with
the microshells, and the hypersurfaces $\{\Sigma_t\}$ of constant
Schwarzschild time (denoted by $t_\infty\equiv t_N$) which are
associated with (distant) static observers~\footnote{Strictly speaking,
the domain of outer communication is well
defined only provided one knows the development of the whole space-time
manifold \cite{pos}.
However, in our case we cannot know from the beginning whether an event
horizon will form or not and, thus, if $\{\Sigma_t\}$ can cover just a
portion or the whole of space-time.
At the same time, $\{\Sigma_\tau\}$ is defined naturally just on the
world-strip where the macroshell has support.}.
Since we shall consider only bound orbits for the shell with
initially large radius and negligible velocity, it is useful, in
this respect, to separate the collapse in the
{\em Newtonian regime\/} (NR) and the {\em near horizon regime\/}
(NHR):
\begin{description}
\item[NR)]
when the radius of the shell is large (and its velocity small),
the difference between the proper time of the microshells and the
Schwarzschild time is negligible, meaning that observers comoving
with microshells at different radii can synchronize their clocks
with the clock of the distant observer.
Therefore, one expects that the Hilbert space which contains
states of the microshells as viewed by the comoving
observers, $\mathcal{H}_\tau=\{\Phi^{(\tau)}(r,\tau)\}$, is the same
as $\mathcal{H}_t=\{\Phi^{(t)}(r,t_\infty)\}$ viewed by the (distant)
static observer.
We examine this case in Appendix~\ref{newton}.
\item[NHR)]
near the gravitational radius, the relative gravitational red-shift
between different parts of the macroshell is no longer negligible
and the two foliations $\{\Sigma_\tau\}$ and $\{\Sigma_t\}$ differ
significantly, since $t_\infty=t_\infty(r,\tau)$ is a large
transformation there [see Eq.~(\ref{times})].
One can still introduce an Hilbert space $\mathcal{H}_\tau$
as relevant for the comoving observers, and a space
$\mathcal{H}_t$ for the distant observer, but the relation
between the two spaces (or their physical equivalence) is far
from trivial, as will be apparent from the corresponding
Schr\"odinger equations we display below and in Appendix~\ref{schwa}
(see, {\em e.g.}, Refs.~\cite{hk,h2} for analogous considerations).
\end{description}
In NR there is no conceptual ambiguity, since the Newtonian
(Schwarzschild) time (equal to the proper time) is a Killing
vector with respect to which canonical quantization is
straightforward and most calculations can then be completed
analytically (see, {\em e.g.}, Appendix~\ref{newton} and
Ref.~\cite{shell}).
Instead, in NHR there is no {\em a priori\/} reason to
prefer either of the two Hilbert spaces to describe properly
the matter in the shell.
Further, if one were to neglect the backreaction on the metric
and study a
shell evolving on a fixed (Schwarzschild) background, the
Schwarzschild time would still be a Killing vector and provide
a unique way for canonical quantization.
However, if one considers a collapsing macroshell with a
microshell structure in order to successively include backreaction,
the canonical analysis is much more involved and ambiguous.
\par
In the spirit of locality, it seems natural (and also appears
much easier) to analyse bound states of microshells with the aid
of $\mathcal{H}_\tau$ and then move on to the point of view of
the static observer by simply making a change of time coordinate
(after having taken the thin shell limit).
However, in so doing one does not expect to recover the space
$\mathcal{H}_t$ obtained prior to the thin shell limit, nor is it
obvious whether the Principle of Equivalence, in any of its forms
\cite{will},
can be trusted for length scales associated with such bound states
which are of the order the Compton wavelength of (elementary)
particles.
In the present paper, we shall motivate taking the thin shell limit
at some point by studying the trajectory of the shell only down to a
radius greater than the gravitational radius (horizon) by several
shell thicknesses.
\par
\begin{figure}
\centerline{\epsfxsize=300pt\epsfbox{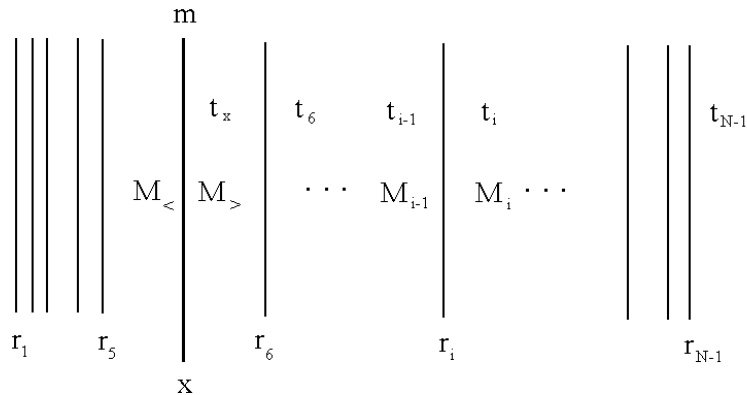}}
\caption{The macroshell structure.}
\label{macro}
\end{figure}

In order to define an effective Hamiltonian for each microshell,
let us single out one of the $N$ microshells, denote its radius by
$x$ and study its motion in the background defined by the remaining
$N-1$ microshells (see Fig.~\ref{macro}).
When $x\not=r_i$, one can apply the junction
conditions \cite{israel} which yield
\be
\left({dx\over d\tau_x}\right)^2=-1+{(M_>-M_<)^2\over m^2}
+G\,{M_>+M_<\over x}+{G^2\,m^2\over 4\,x^2}
\equiv F(x)
\ ,
\label{geo_x}
\ee
where $M_<$ ($M_>\equiv M_x$) is the ADM mass computed on the
inner (outer) surface of the microshell and $\tau_x$ is the
proper time of the microshell.
Then, on multiplying Eq.~(\ref{geo_x}) by $m/2$ one will obtain
the equation of motion for a microshell in the form of an
effective Hamiltonian constraint
\be
H_m\equiv {1\over 2}\,m\,\left({dx\over d\tau_x}\right)^2+V=0
\ .
\label{H_m}
\ee
Since $m\ll M$, we shall take $M_>-M_<\simeq m$, which
amounts to $dx/d\tau_x\simeq 0$ for large $x$ ($\gg R_H$)
and is consistent with the choice of having the microshells
confined within the thickness $\delta$ at large radius.
With this assumption both $M_>$ and $M_<$ become functions
of the microshells distribution $\{r_i,M_i\}$ only, but
the evaluation of $V$ remains extremely involved because
one should first determine the most likely set of $2\,(N-1)$
values $\{r_i,M_i\}$.
\par
One can get an approximate expression for $V$ by making careful
use of the thin shell limit ($\delta\ll r_1$) and considering the form
of the potential when $x<r_1$ or $x>r_1+\delta$.
In this case, if we denote by $X$ the (average) radius of the
macroshell of proper mass $M-m$, we can also supplement
Eq.~(\ref{geo_x}) with the analogous expression for the
macroshell.
\begin{figure}
\centerline{\epsfxsize=300pt\epsfbox{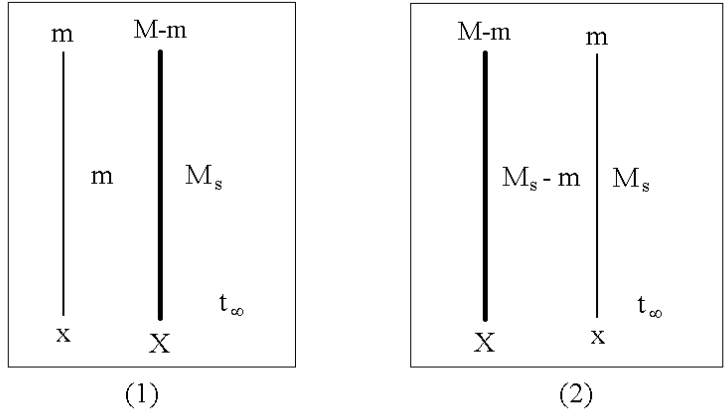}}
\caption{The equivalent configuration when the microshell
lies outside the macroshell: (1) $x<X$; (2) $X<x$.}
\label{ext}
\end{figure}
For $x<X$ [see case (1) in Fig.~\ref{ext}] one then has
\be
&&
\left({dx\over d\tau}\right)^2=
\strut\displaystyle{G\,m\over x}
+{G^2\,m^2\over 4\,x^2}\equiv F_<(x)
\nonumber \\
\label{F<}\\
&&
\left({dX\over d\tau}\right)^2=-1
+\left(\strut\displaystyle{M_s-m\over M-m}\right)^2
+G\,{M_s+m\over X}
+G^2\,{(M-m)^2\over 4\,X^2}
\equiv F_<(X)
\ ,
\nonumber
\ee
where we have equated the proper times of the two shells in
agreement with our choice of foliating the space-time into spatial
slices $\Sigma_\tau$ parameterized by the local time of the
microshells.
Analogously, for $X<x$ [see case (2) in Fig.~\ref{ext}]
\be
&&\left({dx\over d\tau}\right)^2=
G\,\strut\displaystyle{2\,M_s-m\over x}+{G^2\,m^2\over 4\,x^2}
\equiv F_>(x)
\nonumber \\
\label{F>}\\
&&\left({dX\over d\tau}\right)^2
=-1+\left(\strut\displaystyle{M_s-m\over M-m}\right)^2
+G\,{M_s-m\over X}+G^2\,{(M-m)^2\over 4\,X^2}
\equiv F_>(X)
\nonumber
\ .
\ee
One is now ready to define the relative and centre-mass radius
according to
\be
\begin{array}{l}
\bar r\equiv x-X
\\
\\
R\equiv \strut\displaystyle{m\over M}\,x+{\mu\over M}\,X
\ ,
\end{array}
\ee
where $\mu\equiv{m\,(M-m)/M}$ is the reduced mass of the system.
The above relations can then be inverted and, upon substituting
into Eqs.~(\ref{F<}) and (\ref{F>}), one
obtains an effective Hamiltonian for the two shells given by
\be
H^{(\tau)}={1\over 2}\,M\,\left({dR\over d\tau}\right)^2
+{1\over 2}\,\mu\,\left({d\bar r\over d\tau}\right)^2
+V^{(\tau)}\equiv H_M^{(\tau)}+H_m^{(\tau)}
\ .
\ee
The potential contains the following relevant terms
\be
V_m^{(\tau)}=
\strut\displaystyle{G\,M_s\,m\over 2\,R}\,{\bar r\over R}
\times
\left\{\begin{array}{ll}
\left(1-\strut\displaystyle{G\,M^2\over 2\,R\,M_s}
\right)
&
\ \ \bar r>+\strut\displaystyle{\delta\over 2}
\\
&
\\
\left(-1-\strut\displaystyle{G\,M^2\over 2\,R\,M_s}
\right)
&
\ \ \bar r<-\strut\displaystyle{\delta\over 2}\ ,
\end{array}
\right.
\label{linear_tau}
\ee
where we have also assumed $m\ll M_s$ ($\le M$) and neglected all
non-leading terms in $m$.
The assumption $M_s\gg m$, although is not guaranteed
{\em a priori\/}, has been done here just for the purpose
of simplifying the displayed expressions.
However, we shall show in Section~\ref{trajectory} that it
actually holds true all along the collapse in all the cases
of interest.
\par
\begin{figure}
\centerline{\epsfxsize=300pt\epsfbox{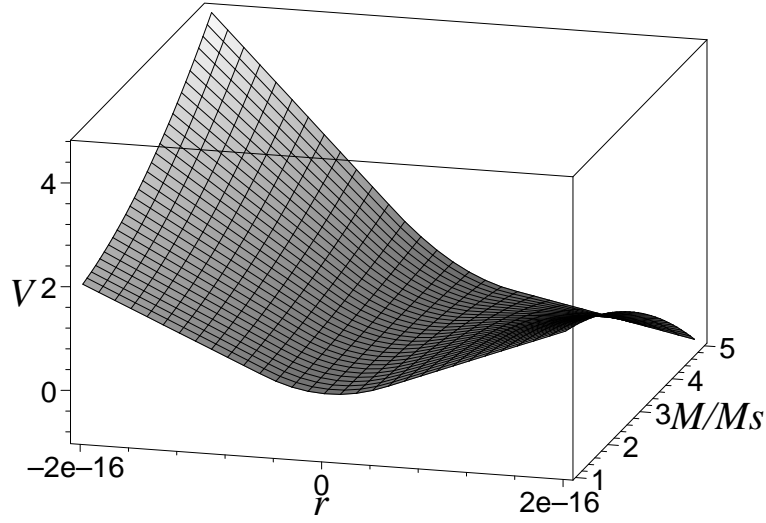}}
\caption{The potential $V^{(\tau)}$ for $M=10^{40}\,m_h$,
$R=4\,R_H=10^{-13}\,$m, $\delta\sim\ell_h=10^{-16}\,$m,
$-\delta<\bar r<\delta$ and $1\le M/M_s\le 5$.
Vertical units are arbitrary.}
\label{Vtau3D_m}
\end{figure}
\begin{figure}
\centerline{\epsfxsize=300pt\epsfbox{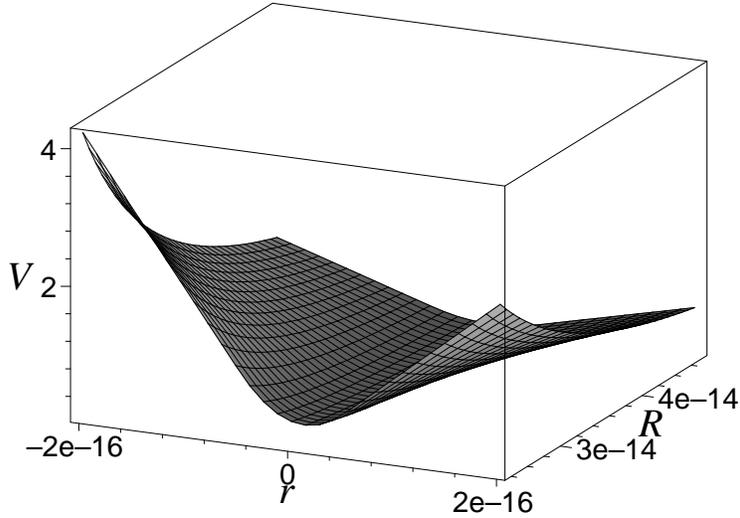}}
\caption{The potential $V^{(\tau)}$ for $M=M_s=10^{40}\,m_h$,
$\delta\sim\ell_h=10^{-16}\,$m, $-\delta<\bar r<\delta$ and
$R_H<R<2\,R_H$.
Vertical units are arbitrary.}
\label{Vtau_r1}
\end{figure}
\begin{figure}
\centerline{\epsfxsize=300pt\epsfbox{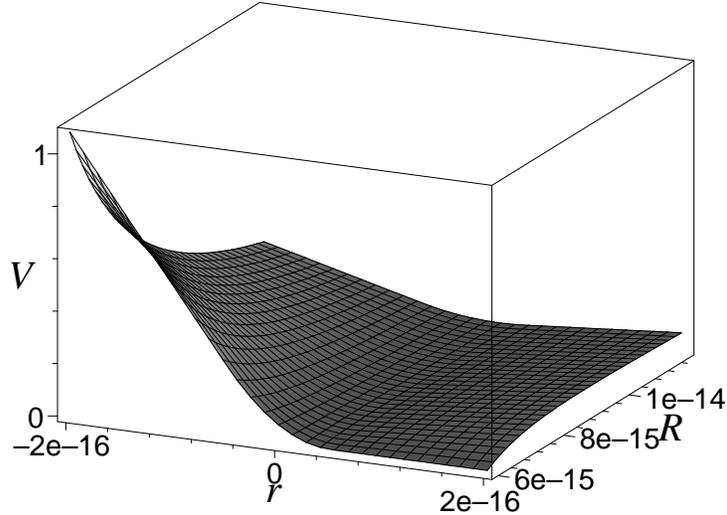}}
\caption{The potential $V^{(\tau)}$ for $M=M_s=10^{40}\,m_h$,
$\delta\sim\ell_h=10^{-16}\,$m, $-\delta<\bar r<\delta$
and $R_H/4<R<R_H/2$.
Vertical units are arbitrary.}
\label{Vtau_r2}
\end{figure}

At this point, one can use the external potential to shape
the distribution of microshells inside the macroshell and
construct the potential $V^{(\tau)}$ for
$|\bar r|<\delta/2$ (see Appendix~\ref{newton} for an explicit
calculation in NR).
By expanding the latter potential in powers of $\delta/R$ and
neglecting terms of order $(\delta/R)^2$ or higher, one finally
obtains an effective Hamiltonian for each microshell,
\be
H_m^{(\tau)}={1\over 2}\,m\,\left({d\bar r\over d\tau}\right)^2
+V_m^{(\tau)}
\ ,
\ee
with a potential function which interpolates smoothly~\footnote{We
assume there is no discontinuity experienced by the test microshell
when it crosses the borders of the macroshell and equate the first
derivative of the quadratic potential to the first derivative of
the linear potential at $|\bar r|=\delta/2$.}
between the two linear in $\bar r$ (outer) parts of the potential
given by $V_m^{(\tau)}$ in Eq.~(\ref{linear_tau}),
\be
V_m^{(\tau)}={G\,M_s\,m\over 2\,R^2}\,
\left({\bar r^2\over \delta}+{\delta\over 4}\right)
-{G^2\,M^2\,m\over 4\,R^3}\,\bar r
\ ,
\ \ \ \ \ \ \ |\bar r|\le {\delta\over 2}
\ .
\label{inner_tau}
\ee
We then observe that for $R\gg R_H\sim 2\,G\,M$ the term
linear in $\bar r$ is negligible and the potential is thus
symmetric around the average radius of the macroshell,
but for $M_s<M$ it becomes steeper for $\bar r<0$ and
flattens out for $\bar r >0$
(see Fig.~\ref{Vtau3D_m} for the case of $N=10^{40}$ protons
and $\delta$ of the order of the Compton wavelength
$\ell_h$ of the proton).
The same effect happens for decreasing values of the average
radius at fixed ratio $M/M_s$ (see Figs.~\ref{Vtau_r1} and
\ref{Vtau_r2}; note that, for this choice of $M$, the potential
confines all the way down to $R_H$).
\par
For a comparison with the case when the space-time is foliated
into slices $\Sigma_t$ parameterized by the Schwarzschild time
$t_\infty$ measured by a (distant) static observer we refer to
Appendix~\ref{schwa}.
There we show how to obtain the relevant potential $V_m^{(t)}$
as an expansion in powers of $G$.
Then, since $V_m^{(t)}$ contains terms of all orders in $G$,
while $V_m^{(\tau)}$ goes only up to $\mathcal{O}(G^2)$, this
comparison clearly confirms that perturbative methods cannot
always be trusted when dealing with strong gravitational
fields or, if we wish, the disadvantage of using the Schwarzschild
rather than proper time.
In the same Appendix we show that, if one takes the this shell
limit of $V_m^{(\tau)}$ and then changes to the Schwarzschild
time multiplying by $(d\tau/dt_\infty)^2$, one recovers the thin
shell limit of $V_m^{(t)}$ in NR.
This means the two steps of introducing a potential and taking
the thin shell limit commute (as expected) in NR.
\par
For $R\gg R_H$ and $M\sim M_s$, the potential $V_m$ is symmetric
around $\bar r=0$.
This confines the microshells within the following thickness
\cite{shell}
(see also Appendix~\ref{schro} for more details on the spectrum
of bound states)
\be
\delta\sim {\ell_m\over (2\,N)^{1/3}}\,
\left({R\over\ell_p}\right)^{2/3}
\sim\ell_m^{2/3}\,R^{1/3}\,\left({R\over R_H}\right)^{1/3}
\ ,
\label{d}
\ee
where $\ell_m=\hbar/m$ is the Compton wavelength of a
microshell.
For $R$ close to $R_H$ or smaller ratios of $M_s/M$ the potential
twists (see Figs.~\ref{Vtau_r2} and \ref{Vtau3D_m})
and, eventually, ceases to confine the microshells on the outside
if the macroshell shrinks below the gravitational radius
[see Eq.~(\ref{deco})].
We further note that the spreads of the lowest wavefunctions
are also of order $\delta$, thus the bosonic microshells
are essentially superimposed and form a condensate,
hence some classical singular behaviour, such as microshell
crossings, do not occur.
\par
Let us now summarize what we have done in simple quantum mechamical
terms.
Since we are considering wavefunctions, we have viewed each microshell
as immersed in a continuous matter distribution, corresponding to the
mean field of the others which are then treated as a (wide) single
shell (Hartree approximation or independent particle model - see also
the Thomas-Fermi model of the atom \cite{messiah,shell}).
Then we noted that, inside the macroshell, a microshell can
undergo small displacements ($\delta/R$ si small) about equilibrium,
which will be harmonic, and on requiring continuity with the
potential outside the shell one obtains Eq.~(\ref{inner_tau}).
Further, since in the context of our Hartree approximation the total
wavefunction is just the product of the individual
ones~\footnote{Let us also remember that the many-boson
wavefunction is symmetric.}, it will be sufficient in what follows
to just exhibit the Lagrangian and Hamiltonian for a given microshell.
\section{The radiation and backreaction}
\setcounter{equation}{0}
\label{f&t}
In this Section we use the description of the macroshell as given in
Section~\ref{structure} and add the coupling to an external radiation
field.
This will allow the shell to radiate away any excess of the proper energy
induced by the non-adiabaticity of the collapse, as outlined in the
Introduction.
\par
Our aim is to evaluate the flux of emitted radiation, which will determine
$M_s=M_s(\tau)$, and the corresponding backreaction on the trajectory
$R=R(\tau)$.
We shall first estimate the non-adiabatic amplitude of excitation
and, subsequently, give a purely quantum mechanical (coherent) treatment
of the entire process of excitation and emission by coupling explicitly
the microshell modes $\Phi_n$ to the radiation field.
The outcome will be a set of two coupled ordinary differential equations
(for $R$ and $M_s$) which we shall solve numerically.
\subsection{The excitation amplitude}
\label{s_fqp}
From the previous Section we know that the microshells are governed
by the Schr\"odinger equation (see also Appendix~\ref{schro})
\be
i\,\hbar\,{\partial \Phi_s\over\partial \tau} =
\hat H_m^{(\tau)}\,\Phi_s
\label{Schrod}
\ee
with
\be
\hat H_m^{(\tau)}={\hat\pi_r^2\over 2\,m}+V^{(\tau)}_m
\ ,
\ee
where $V_m^{(\tau)}$ is given in Eq.~(\ref{inner_tau}) and has an
explicit time dependence due to $R=R(\tau)$.
This type of equation can be solved by making use of invariant
operators $\hat I=\hat I(\tau)$ which satisfy \cite{lewis-ries}
\be
i\,\hbar\,{\partial \hat I\over\partial\tau} +
\left[\hat I,\hat H_m^{(\tau)}\right] =0
\ .
\ee
The general solution $\Phi_s=\Phi_s(\tau)$ can then be
written in the form
\be
\ket{\Phi,\,\tau}_{Is}=
\sum_n c_n\,e^{i\,\varphi_n}\,\ket{n,\tau}_I
\ ,
\ee
where $\ket{n,\tau}_I$ is an eigenvector of $\hat I$
with time-independent eigenvalue $\lambda_n$ and the $c_n$
are complex coefficients.
We also recall that the phase $\varphi_n=\varphi_n(\tau)$
is given by the sum of the geometrical and dynamical phases,
\be
\varphi_n={i\over\hbar}\,\int_{\tau_0}^{\tau}
\ _I\,\bra{n,\tau'}\,\hbar\,\partial_{\tau'} +
i\,\hat H_m^{(\tau)}\,\ket{n,\tau'}_I\,d\tau'
\ .
\ee
\par
The Hamiltonian in Eq.~(\ref{Schrod}) describes a harmonic
oscillator of fixed mass $m$ and variable frequency
\be
\Omega={1\over R}\,\sqrt{{R_H\over 2\,\delta}}
\ .
\label{Omega}
\ee
Hence, one can introduce standard annihilation and creation
operators $\a$ and $\ac$,
\be
\a&=&
\sqrt{m\,\Omega\over 2\,\hbar}\,
\left(\bar r +{i\,\hat \pi_r\over m\,\Omega}\right)
\ ,
\ee
with $\left[\a,\ac\right]=1$ and define the vacuum state as
\be
\a\,\ket{0,\tau}_a=0
\ee
A complete set of eigenstates
${\cal A}=\left\{\ket{n,\tau}_a \right\}$
is then given by
\be
\ket{n,\tau}_a=
{\left(\ac\right)^n\over\sqrt{n!}}\,\ket{0,\tau}_a
\ ,
\ee
where
\be
&&\a\,\ket{n,\tau}_a=\sqrt{n}\,\ket{n-1,\tau}_a
\nonumber \\
\\
&&\ac\,\ket{n,\tau}_a=\sqrt{n+1}\,\ket{n+1,\tau}_a
\ .
\nonumber
\ee
The solutions of the corresponding time-independent problem
obtained by assuming $R$ constant and displayed in Appendix~\ref{schro}
are here recovered as $\Phi_n^{(2)}(\bar r)=\pro{\bar r}{n,0}_a$.
\par
One can also introduce  the linear (non-Hermitian) annihilation
and creation invariants $\b$ and $\bc$ \cite{lewis-ries,gao},
\be
\b={1\over{\sqrt{2\,\hbar}}}\,\left[
{\bar r\over x}+i\,\left(x\,\hat\pi_r-m\,\dot x\,\bar r\right)
\right]
\ ,
\ee
where $\left[\b,\bc\right]=1$ and the function $x=x(\tau)$ is a
solution of~\footnote{A dot denotes derivative with respect
to $\tau$ throughout the paper.}
\be
\ddot{x}+\Omega^2\,x={1\over m^2\,x^3}
\ ,
\label{x}
\ee
with suitable initial conditions.
The system then admits an invariant ground state defined by
\be
\b\,\ket{0,\tau}_b =0
\ ,
\ee
from which one can build a basis of invariant eigenstates
${\cal B}=\{\ket{n,\tau}_b\}$,
\be
\ket{n,\tau}_{bs}&\equiv&
e^{i\,\varphi_n}\,{(\hat b^{\dagger})^n\over \sqrt{n!}}
\,\ket{0,\tau}_b=e^{i\,\varphi_n}\,\ket{n,\tau}_b
\ ,
\label{n-states}
\ee
where
\be
&&\b\,\ket{ n, \tau }_b = \sqrt{n}\,\ket{ n-1,\tau}_b
\nonumber \\
\\
&&\bc\,\ket{ n, \tau }_b = \sqrt{n+1}\,\ket{ n+1,\tau}_b
\ .
\nonumber
\ee
We can then introduce the invariant number operator (in analogy
with the standard number operator $\ac\,\a$)
\be
\hat I_c=\bc\,\b
\ .
\ee
\par
The two basis ${\cal A}$ and ${\cal B} $ are related by Bogoliubov
coefficients according to
\be
\begin{array}{lcr}
\left\{
\begin{array}{l}
\a=B^*\,\b+A^*\,\bc
\\
\\
\ac=B\,\bc+A\,\b
\end{array}\right.
&\Leftrightarrow&
\left\{
\begin{array}{l}
\b=B\,\a-A^*\,\ac
\\
\\
\bc=B^*\,\ac-A\,\a
\ ,
\end{array}
\right.
\end{array}
\label{Bogoliubov}
\ee
where
\be
&&A={1\over 2}\,\sqrt{m\,\Omega}\,\left[\,
\left(x-{1\over m\,\Omega\,x}\right)
-i\,{\dot x\over\Omega}\right]
\nonumber \\
\\
&&B={1\over 2}\,\sqrt{m\,\Omega}\,\left[
\left(x+{1\over m\,\Omega\,x}\right)
-i\,{\dot x\over\Omega}\right]
\ .
\nonumber
\ee
Further, ${\cal A}$ and ${\cal B} $ will coincide at $\tau=0$
if
\be
\b(0)=\a(0)\ \ \ \Rightarrow\ \ \
\hbar\,\Omega\,\left(\hat I _c(0)+{1\over 2}\right)
=\hat H_m(0)
\ ,
\ee
that is $A(0)=0$ and $B(0)=1$, which means that for the function
$x$ we must require
\be
\left\{\begin{array}{l}
x(0)=\strut\displaystyle{1\over\sqrt{m\,\Omega(0)}}
\\
\\
\dot x(0)=0
\ .
\end{array}\right.
\label{init}
\ee
\par
Since $\dot\Omega\sim \dot R$ and $\dot R(0)\simeq 0$, an approximate
solution to Eq.~(\ref{x}) compatible with the initial conditions in
Eq.~(\ref{init}) is given by
\be
x\simeq {1\over\sqrt{m\,\Omega}}
\ \ \ \Rightarrow\ \ \
\dot x\simeq-{\dot\Omega\over2\,\sqrt{m\,\Omega^3}}
\ ,
\ee
provided $\tau$ is sufficiently short so that $\ddot x$ is negligible.
The corresponding Bogoliubov coefficients are then given by
\be
A\simeq-{i\,\dot\Omega\over4\,\Omega^2}
\ , \ \ \ \ B\simeq 1+A
\ ,
\ee
and satisfy the unitarity condition
\be
|B|^2-|A|^2=1
\ .
\ee
\par
The amplitude of the transition from the state $\Phi_0$ to
$\Phi_n$ after the time $\tau\ge 0$ can be easily computed.
In particular, one finds \cite{lewis-ries}
\be
A_{0\to 2\,n}(\tau)&=&\,_a\pro{2\,n,\tau}{0,\tau}_b
\nonumber \\
&=&{1\over \sqrt{B}}\,
\left({A^*\over B}\right)^{n}\,{\sqrt{(2\,n)!}\over 2^n\,n!}
\\
\nonumber \\
A_{0\to 2\,n+1}(\tau)&=&\,_a\pro{2\,n+1,\tau}{0,\tau}_b
\nonumber \\
&=&0
\ .
\ee
To leading order in $|A|$ ({\em i.e.}, $\dot\Omega\sim\dot R$), one
then has $A_{0\to 0}(\tau)\simeq 1$ and the amplitude
of the transition to the excited state with energy
$E_{2\,n}=E_0+2\,n\,\hbar\,\Omega$ is
\be
A_{0\to 2\,n}(\tau)
&\simeq&i^n\,{\sqrt{(2\,n)!}\over 2^{3\,n}\,n!}\,
{\dot\Omega^n(\tau)\over\Omega^{2n}(\tau)}
\nonumber \\
&=&(-i)^n\,{\sqrt{(2\,n)!}\over 3^{n}\,2^{n/2}\,n!}\,
\left({\delta\over R_H}\right)^{n/2}\,\dot R^{n}(\tau)
\ ,
\label{0_2n}
\ee
where we have assumed $R_H=2\,G\,M_s$ remains constant in the
interval $(0,\tau)$.
The relevant expression for us is thus given by Eq.~(\ref{0_2n}), in
which we recall that $\dot R<0$.
We also remark that the above transition amplitude follows as both a
non-adiabatic effect (since $A_{0\to 2\,n}\propto\dot R^{n}$) and a
consequence of the finite thickness of the macroshell
(since $A_{0\to 2\,n}\propto \delta^{n/2}$), the latter being further
related to the quantum mechanical nature of the bound states
[since $\delta\propto\hbar^{2/3}$].
\par
An order of magnitude estimate for the most probable transition
($2\,n=2$) can be obtained by setting $\dot R^2\sim G\,M_s/R$ and
using Eq.~(\ref{d}), in which case
\be
|A_{0\to 2}(R,N)|^2\le |A_{0\to 2}(R_H)|^2\sim
\left({\ell_m\over \ell_p}\right)^{4/3}\,{1\over N^{2/3}}
\ ,
\ee
where we also approximated $M_s$ as $N\,m$ in order to obtain
a function of just the number $N$ of microshells.
Further, one can also check that the above probability is
less than one and, for realistic cases, quite small (we shall
exhibit this later).
This implies that the probability for a microshell to get excited
a second time (after it has radiated away the energy $E_{2}$ once,
see next Section) is negligible and, considering also that the
emission probability is small, we shall not consider double
emission.
We further note that, as expected, the larger $R_H$ the smaller
the ``tidal'' effects and the excitation amplitude.
\subsection{The (thermal) radiation}
\label{radiation}
The problem of coupling a radiation field in the form of a
conformal scalar field to collapsing matter was previously
treated in Ref.~\cite{cv} (see also Ref.~\cite{qg96}) for the
case of a collapsing sphere of dust, for which we recovered
the Hawking temperature in a suitable approximation.
For the present situation, we shall find it more convenient to
use the analogy with an accelerated observer in flat space-time
as put forward in Ref.~\cite{oss2}.
\par
The basic observation is that, if the proper mass or the ADM
mass (or both) vary with time, the collapse of the shell is not
a free fall.
This can be seen very easily by noting that the solutions to
the equation of motion (\ref{geo_x}) when $M_>$ and/or $m$
depend on time differ from the case when the same parameters
are constant.
Alternatively, one can use the second junction condition
\cite{israel} to compute the surface tension ${\cal P}$ of the
shell.
As we have shown in the previous Section, the proper mass $M$
of the shell increases in time, therefore, although our choice
of initial conditions for $R$, $M$ and $M_s$ are such that
${\cal P}(0)=0$, the tension subsequently increases when $M$
increases.
The tension then slows down the collapse \cite{shell}, and this
effect should be further sustained by the emission \cite{vaidya},
which very quickly brings the proper mass back to the initial
value.
As we outlined in the Introduction, we shall then consider $M$
as effectively constant along the collapse and compute the net
variation of $M_s$ in time.
\par
Let us denote by
\be
a=\ddot R-\ddot R_{\rm free}
\ ,
\ee
the difference between the actual acceleration $\ddot R$ of the
shell and that of a freely falling body, $\ddot R_{\rm free}$,
having the same radial position $R$ and proper velocity $\dot R$.
Of course, $a$ would equal the surface gravity,
\be
a_H={1\over4\,G\,M_s}
\ ,
\label{aH}
\ee
of a black hole of mass $M_s$ for $R=R_H$ constant.
In this limiting case, one could exploit the analogy between
Rindler coordinates for an accelerated observer in flat
space-time and Schwarzschild coordinates for static observers
in a black hole background \cite{birrell} and find that the
shell emits the excess energy with the Hawking temperature
\be
T_H={\hbar\,a_H\over 2\,\pi\,k_B}
\ ,
\ee
where $k_B$ is the Boltzmann constant (an explicit construction
which leads to this result was proposed in Ref.~\cite{cv}).
In general, however, the effective acceleration $a$ will have
no {\em a priori\/} fixed relation with $M_s$ and
one should keep it as an independent function of $\tau$ to be
determined later consistently with the equation of motion
of the shell.
It is also clear that, should $\dot R$ turn out to be constant,
$a$ will correspond to the acceleration associated with the
instantaneous ADM mass and distance from the horizon.
\par
We then consider an isotropic massless scalar field $\varphi$
conformally coupled to gravity \cite{birrell} and to the microshells.
The Lagrangian density for a given microshell and the radiation field
is given by
\be
{\mathcal L}&=&
+{1\over r^2}\,
\sum_{n\ge 0}\,\left[{i\,\hbar\over 2}\,\left(
\Phi_{n}^*\,\dot\Phi_{n}^{\phantom{*}}
-\dot\Phi_{n}^*\,\Phi_{n}^{\phantom{*}}
\right)
-{\hbar^2\over 2\,m}\,{\partial\Phi_{n}^*\over \partial r}\,
{\partial\Phi_{n}^{\phantom{*}}\over \partial r}
-V_{m}^{(\tau)}\,\Phi_{n}^*\,\Phi_{n}^{\phantom{*}}
-e\,\sum_{l\not=n}\,
\Phi_{n}^*\,\Phi_{l}^{\phantom{*}}\,
\varphi\right]
\nonumber \\
&&-\left[\partial_\mu\varphi\,\partial^\mu\varphi
+{1\over 6}\,{\cal R}\,\varphi^2\right]
\ ,
\label{L}
\ee
where ${\cal R}$ is the curvature scalar and $V_m^{(\tau)}$ is
the potential given by Eqs.~(\ref{linear_tau}) and
(\ref{inner_tau}).
The last term in the first square bracket represents minus
the interacting Hamiltonian $H_{\rm int}$ with $e$ the coupling
constant of the radiation to a microshell while the wavefunctions
of the latter are expressed in terms of the states found in
Appendix~\ref{schro} as $\Phi_{n}/r$, where we exhibit the energy
levels ($n\ge 0$) so that the form of the interaction can be chosen
in such a way that -- to lowest order -- it just contributes to the
transition between different levels and does not modify the macroshell
ground state energy.
Further, the factor of $(1/r)$ comes from the normalization measure
now being $r^2\,dr$.
In fact, in the thin shell limit it is consistent to approximate the
metric on the shell with the outer Vaidya line element
\cite{vaidya1,vaidya2}.
Let us again note that we have just exibited the Lagrangian for
a given microshell since in the Hartree approximation the total
wavefunction is just the product of the individual
ones~\footnote{Of course, both the total and the single particle
wavefunctions are normalized to unity.}.
\par
Another important observation comes into play at this point.
In the preceding Section we have computed the (non-vanishing)
transition amplitudes $A_{0\to 2n}$ to lowest order in $\dot R$,
that is ${\cal O}(\dot R)$ [see Eq.~(\ref{0_2n})], with $M$ and
$M_s$ held constant.
Therefore, in order to determine the flux of radiation to lowest
order in $\dot R$, it is sufficient to compute the
probability of emission to ${\cal O}(\dot R^0)$ with $M$ and $M_s$
constant.
In this approximation ($M_s$ constant), the Vaidya metric reduces
to the Schwarzschild metric, so that the relevant four-dimensional
measure of integration is
\be
\sqrt{-g}\,d^4x\simeq
r^2\,\sin^2\theta\,d\theta\,d\phi\,dr\,dt
\ ,
\ee
where $t$ is the Schwarzschild time measured by a static observer
(denoted by $t_\infty$ in previous Sections),
\be
t\sim {\tau\over\sqrt{1-R_H/R}}
\ .
\label{ts}
\ee
Finally, since the functions $\Phi_{n}$ are spherically symmetric
and peaked near $r=R(\tau)$, it is convenient to integrate over
the angular coordinates and write
\be
\sqrt{-g}\,d^4x=4\,\pi\,(R+\bar r)^2\,d\bar r\,dt
\ ,
\ee
where $\bar r$ is the relative radial coordinate and $R=R(t)$
is the trajectory of the macroshell as before.
\par
After the lapse of proper time $\tau$, each microshell will jump
into the excited state of energy $E_{2n}=E_0+2\,n\,\hbar\,\Omega$
with a transition amplitude $A_{0\to 2n}(\tau)$.
Successively, it can decay back to the ground state by emitting
a quantum of energy $\hbar\,\omega$ (having a wavelength much
larger than the shell thickness, see Table~\ref{table1})
of the scalar field.
We shall later see that the macroshell velocity is small and
the emissions are numerous.
As a consequence, the distance covered between each emission is
small with respect to the wavelength and therefore the emitted
radiation will be coherent.
\par
Let us now estimate, using first order perturbation theory
in $e$, the transition amplitude for the emission of quanta
of the scalar field, which will occur when the microshells in the
state $\Phi_{2n}$ decay back to the ground state $\Phi_0$, for
the interval between the times
$t_1\equiv t(\tau_1)$ and $t_2\equiv t(\tau_2)$
($0<\tau_1<\tau_2$).
It will be given by
\be
-{i\over\hbar}\,\int_{t_1}^{t_2} dt\,\bra{\rm final}\,
\hat H_{\rm int}\,\ket{\rm initial}
&=&-{i\,e\over\hbar}\,\sum_{i=1}^N\,
\int_{t_1}^{t_2} dt\,
\bra{1_{\omega};t}\,\hat\varphi\,\ket{0;t_1}\,
\prod_{j=1}^N\,\bra{j,0;t}\,
\nonumber \\
&&
\times
\sum_{n\ge 0}\,\sum_{l\not= n}\,
\hat\Phi_{i,n}^*\,\hat\Phi_{i,l}^{\phantom{*}}\,
\sum_{s\ge 0}\,\prod_{r=1}^N\,\ket{r,s;t}\,
\pro{r,s;t}{r,0;0}
\nonumber \\
&\simeq&
-{4\,\pi\,i\,e\over\hbar}\,
\sum_{i=1}^N\,
\int_{t_1}^{t_2} dt\,
\int d\bar r_i\,
\Phi_{0}(\bar r_i)\,\Phi_{2}(\bar r_i)\,A_{i,0\to 2}(t,0)
\nonumber \\
&&
\times
e^{-2\,i\,\Omega\,\tau}\,
\bra{1_{\omega}}\,\hat\varphi(R+\bar r_i,t)\,\ket{0}
\ ,
\label{ta}
\ee
where $\pro{i,n;t}{i,0;0}=A_{i,0\to n}(t)$ is the excitation
amplitude (for the i$^{th}$ microshell) which, as computed in the
previous Section, is dominated by the contribution with $n=2$ and
for the photon state $\ket{0}$ we consider the Unruh
vacuum, corresponding to an outgoing flux of radiation
\cite{birrell}.
This of course implies that $a$ is non-zero, indeed
we shall actually find that $a\sim -\ddot R_{\rm free}$
(since $\ddot R\simeq 0$), consistently with our choice of vacuum and,
near the horizon, the flux corresponds to Hawking radiation
(see Ref.~\cite{brout} for a detailed discussion of the
analogy between Schwarzschild and Rindler spaces).
Let us further note that, although a black hole is only close
to being formed, it is generally accepted that it is the Unruh
vacuum that best approximates the state that would be formed
following the collapse of a massive body~\cite{birrell}.
\par
On taking the modulus squared of Eq.~(\ref{ta}), one obtains the
transition probability
\be
P(2\,\Omega;t_2,t_1)&\simeq&
{16\,\pi^2\,e^2\over \hbar^2}\,
\sum_{i=1}^N\,\sum_{j=1}^N\,
\int_{t_1}^{t_2}dt''\,\int_{t_1}^{t_2}dt'\,
\int d\bar r''_i\,\int d\bar r'_j\,
A''_{i,0\to 2}\,A'_{j,0\to 2}\,
e^{-i\,2\,(\Omega''\,\tau''-\Omega'\,\tau')}
\nonumber \\
&&\times
\Phi_0(\bar r''_i)\,\Phi_2(\bar r''_i)\,
\Phi_2(\bar r'_j)\,\Phi_0(\bar r'_j)\,
\bra{0}\,\hat\varphi(R''+\bar r''_i,t'')\,
\hat\varphi(R'+\bar r'_j,t')\,\ket{0}
\ ,
\label{p}
\ee
where $X'\equiv X(t')$, $X''\equiv X(t'')$ for any function of
time and the last term is the Wightman function for the scalar
field $\varphi$.
\par
The four-dimensional Wightman function can be related to its
two-dimensional counterpart $D_u^+$ \cite{birrell} through
\be
\bra{0}\,\hat\varphi(R''+\bar r''_i,t'')\,
\hat\varphi(R'+\bar r'_j,t')\,\ket{0}
&=&{D_u^+(R''+\bar r''_i,t'';R'+\bar r'_j,t')\over 4\,\pi\,
(R''+\bar r''_i)\,(R'+\bar r'_j)}
\nonumber \\
&=&
-{\hbar\over 16\,\pi^2}\,
{\ln\left[(\Delta v_{ij}-i\,\epsilon)\,
(\Delta\bar u_{ij}-i\,\epsilon)\right]
\over (R''+\bar r''_i)\,(R'+\bar r'_j)}
\ ,
\ee
where, of course, we only consider $s$-waves and
\be
\Delta v_{ij}&=&t''-t'+(R''+\bar r''_i)_*-(R'+\bar r'_j)_*
\nonumber \\
&\simeq&
\Delta t+{\bar R\,(\Delta R+\Delta\bar r_{ij})\over \bar R-R_H}
\ .
\label{dv}
\ee
In the above $r_*=r+R_H\,\ln[(r/R_H)-1]$ is the turtle coordinate
\cite{mtw}.
We also set $\Delta t=t''-t'$
(similarly for $\Delta R$ and $\Delta\bar r$),
$\bar R=(R''+R')/2$ and assumed $\bar R-R_H\gg|\Delta R|$,
$|\Delta\bar r_{ij}|=|\bar r_i-\bar r_j|$.
Further,
\be
\Delta\bar u_{ij}&=&
-2\,R_H\,\left\{
\exp\left[-{t''-(R''+\bar r''_i)_*\over 2\,R_H}\right]
-\exp\left[-{t'-(R'+\bar r'_j)_*\over 2\,R_H}\right]\right\}
\nonumber \\
&\simeq&
2\,R_H\,\exp\left({\bar R_*-T-t_1\over 2\,R_H}\right)\,
\left\{
\exp\left[{1\over 2\,R_H}\,\left({\Delta t\over 2}
-{\bar R\,(\bar r'_i+\Delta R/2)\over \bar R-R_H}
\right)\right]
\right.
\nonumber \\
&&
\phantom{2\,R_H\,\exp\left\{{\bar R_*-T-2\,t_1\over 2\,R_H}\right\}
\left[\right.}
\left.
-\exp\left[{1\over 2\,R_H}\,\left(
{\bar R\,(\bar r''_j+\Delta R/2)\over \bar R-R_H}
-{\Delta t\over 2}\right)\right]
\right\}
\ ,
\label{du}
\ee
where $T=(t'+t'')/2-t_1$ and we have performed the same approximation
as we used in obtaining Eq.~(\ref{dv}).
\par
The above expressions can be simplified on noting that one expects
$\Delta t\gg\delta$, $\Delta R$ ($=\dot R\,\Delta t\,\sqrt{1-R_H/R}$)
corresponding to the fact that the wavelength of the emitted scalar
quantum is much greater than the width of the shell and the velocity
of the shell sufficiently small so that it is not displaced much
between one emission and the next (see Table~\ref{table1}).
One then obtains for Eqs.~(\ref{dv}) and (\ref{du}) respectively
\be
\Delta v\simeq\Delta t
\label{dv1}
\ee
\be
\Delta\bar u&\simeq&
2\,R_H\,\exp\left({\bar R_*-T-t_1\over 2\,R_H}\right)\,
\left[
\exp\left({\Delta t\over 4\,R_H}\right)
-\exp\left(-{\Delta t\over 4\,R_H}\right)
\right]
\nonumber \\
&=&
4\,R_H\,\exp\left({\bar R_*-T-t_1\over 2\,R_H}\right)\,
\sinh\left({\Delta t\over 4\,R_H}\right)
\ .
\label{du1}
\ee
\par
On introducing $L=t_2-t_1$, expressing Eq.~(\ref{p}) in terms
of $T$ and $\Delta t$ and approximating the integration measure
as in Eqs.~(\ref{dv1}) and (\ref{du1}) with
\be
&&R'=R(T)-\dot R(T)\,\sqrt{1-R_H/R}\,(\Delta t/2)
\nonumber \\
\\
&&R''=R(T)+\dot R(T)\,\sqrt{1-R_H/R}\,(\Delta t/2)
\ ,
\nonumber
\ee
and $R(T)=\bar R$, one obtains to leading order in $\dot R$
\be
P(2\,\Omega;t_1,L)&\simeq&
N^2\,{e^2\,\hbar\over 2\,m^2}\,
\int_0^L dT\,{|A_{0\to 2}|^2\over R^6\,\Omega^2}\,
\int_{-\Lambda}^{+\Lambda}
d(\Delta t)\,
e^{-i\,2\,\Omega\,\sqrt{1-R_H/R}\,\Delta t}
\nonumber \\
&&\times\ln\left[
4\,R_H\,e^{(R_*-T-t_1)/2\,R_H}\,
(\Delta t-i\,\epsilon)\,
\sinh\left({\Delta t\over 4\,R_H}-i\,\epsilon\right)\right]
\nonumber \\
&\simeq&
N^2\,{e^2\,\hbar\over 2\,m^2}\,
\int_0^L dT\,
{|A_{0\to 2}|^2\over R^6\,\Omega^2}\,\int_{-\Lambda}^{+\Lambda}
d(\Delta t)\,
e^{-i\,2\,\Omega\,\sqrt{1-R_H/R}\,\Delta t}
\nonumber \\
&&\ \
\times
{(-i)\over2\,\Omega\,\sqrt{1-R_H/R}}\,
\left[{1\over\Delta t-i\,\epsilon}
+{1\over 4\,R_H}\,
\coth\left({\Delta t\over 4\,R_H}-i\,\epsilon\right)\right]
\ ,
\label{p1}
\ee
with $\Lambda=L-2\,|T-L/2|$.
In the above, we approximated
$\tau''-\tau'\simeq \Delta t\,\sqrt{1-R_H/R}$ according to
Eqs.~(\ref{ts}) and the last equality follows from integration
by parts in $\Delta t$ (neglecting boundary terms).
\par
The integral (\ref{p1}) can be computed by analytic continuation
in the complex $\Delta t$ plane in which, upon using the formula
\be
\coth(\pi\,x)={1\over\pi\,x}
-i\,{x\over\pi}\,\sum_{0\not=n=-\infty}^{+\infty}
\,{1\over n\,(x-i\,n)}
\ ,
\ee
one finds the poles
\be
\Delta t_n=4\,\pi\,n\,i\,R_H
\ ,
\ee
for any integer value of $n$.
For $\Omega>0$ (corresponding to emission) the contour of
integration can be closed along an arc in the lower half plane
and we get
\be
P(2\,\Omega;t_1,L)&\simeq&
N^2\,\,
{\pi\,e^2\,\hbar\over 2\,m^2}\,
\int_0^L {dT\,|A_{0\to 2}|^2\over R^6\,\Omega^3\,\sqrt{1-R_H/R}}\,
\sum_{n=1}^{N_L}\,e^{-8\,n\,\pi\,\Omega\,R_H\,\sqrt{1-R_H/R}}
\nonumber \\
&&
+P_L
\ ,
\label{p2}
\ee
where $N_L$ is the maximum value of $n$ for which the pole
$\Delta t_n$ lies within the contour of integration and $P_L$
is the contribution from the arc.
\par
If $N_L$ is sufficiently large, then $P_L$ can be neglected,
\be
\sum_{n=1}^{N_L}\,
e^{-8\,n\,\pi\,\Omega\,R_H\,\sqrt{1-R_H/R}}
\sim {1\over e^{2\,\hbar\,\Omega/k_B\,{\cal T}}-1}
\ ,
\label{planck}
\ee
and one recovers the Planck distribution with (instantaneous)
temperature
\be
{\cal T}={T_H\over\sqrt{1-R_H/R}}
\ .
\label{T}
\ee
In general, however, one expects that the time $L$ is not long
enough and non-thermal contributions will render the evaluation of
the probability of emission very complicated.
One should then divide the time of collapse into small
intervals and compute all the relevant quantities step by step.
This procedure can be simplified by employing an approximation
introduced in order to follow a trajectory \cite{oss2}
and which amounts to estimating the probability of emission per
unit time at a given value of $t$ (or $\tau$) as
\be
{dP(\Omega;t)\over dL}\sim
{\pi\,e^2\,\hbar\over 2\,m^2}\,
{N^2\,|A_{0\to 2}|^2\,\over R^6\,\Omega^3\,\sqrt{1-R_H/R}}\,
{1\over e^{2\,\hbar\,\Omega/k_B\,{\cal T}}-1}
\ ,
\label{ave}
\ee
in which all time-dependent quantities must be evaluated at the same
time $t$ and the right hand side is to lowest order in $\dot R$.
\par
We finally recall that in Ref.~\cite{cv} we proposed that the finite
thickness of the macroshell could provide a solution to the problem
of ultra-Planckian frequencies.
In fact, if $\omega^*$ is the frequency of the emitted quanta as is
measured by a distant observer, a fixed observer located near the
point of emission at $r=R$ will instead measure a blue-shifted
frequency
\be
\omega=\left(1-{R_H\over R}\right)^{-1/2}\,\omega^*
\ ,
\ee
and this expression clearly gives $\omega>1/\ell_p$
for $R$ sufficiently close to $R_H$.
In order to probe these modes, one should use a detector
localized in a region smaller than $\omega^{-1}\sim\ell_p$,
or, alternatively, one expects that the spread $\delta$ of the
wavefunctions $\Phi_n$ should be less than $\ell_p$ for our
collapsing microshells to couple with conformal quanta of
ultra-Planckian energies.
However, as we have shown in Refs.~\cite{oss,cv}, taking
$\delta/\ell_p\to 0$ decouples the emitter from the radiation and
ultra-Planckian frequencies are not excited (for a similar argument
against ultra-Planckian frequencies, see Ref.~\cite{helfer}).
\subsection{The flux}
\label{flux}
It is now straightforward to estimate the total flux of emitted
radiation (the {\em luminosity}) as a function of time.
In particular, the rate of proper energy lost by the macroshell
per unit (Schwarzschild) time of a static observer placed at large
$r$ is given by
\be
{dE\over dt}=-{\cal A}\,
\sum_\omega\,\mu(\omega)\,\Gamma(\omega)\,\hbar\,\omega\,
{dP(\omega;t)\over dL}
\ ,
\label{lum}
\ee
where ${\cal A}=4\,\pi\,R^2$ is the surface area of the shell,
$\mu(\omega)=(1-R_H/R)^{3/2}\,\omega^2$ the phase space measure
for photons of frequency $\omega$ (measured at the shell
position $r=R$),
$\Gamma\sim 1$ the grey-body factor for zero angular momentum
outgoing scalar waves \cite{page} and the last factor is obtained
from the approximation (\ref{ave}).
The sum in Eq.~(\ref{lum}) is dominated by the contribution with
$\omega=2\,\Omega$ and one obtains
\be
{dE\over dt}\simeq
-{16\,\pi^2\,e^2\over 9}\,{N^2\,\ell_m^2\over R^4}\,
\left({\delta\over R_H}\right)\,
\left(1-{R_H\over R}\right)\,
{\dot R^2\over e^{2\,\Omega\,\sqrt{1-R_H/R}/k_B\,T_H}-1}
\ .
\ee
We can now substitute the expressions (\ref{Omega}) for $\Omega$ and
(\ref{d}) for $\delta$ and (minus) the flux becomes
\be
{dE\over dt}\simeq
-{16\,\pi^2\,e^2\over 9}\,
{N^2\,\ell_m^{8/3}\over R_{\phantom H}^{10/3}\,R_H^{4/3}}\,
\left(1-{R_H\over R}\right)\,
{\dot R^2\over e^{2\,\Omega\,\sqrt{1-R_H/R}/k_B\,T_H}-1}
\ .
\ee
\par
We note that, for $R$ close to $R_H$ the Boltzmann exponent is
very small (because of the Tolman factor).
Upon expanding the exponent and keeping the leading order in $R-R_H$
we obtain
\be
{dE\over dt}\sim
-e^2\,{N^2\over R^2}\,
\left({\ell_m\over R_H}\right)^3\,
\left(1-{R_H\over R}\right)^{1/2}\,\dot R^2
\ .
\ee
The flux should therefore vanish for $R\to R_H^+$ and, on comparing
with the general relation given in Ref.~\cite{vaidya},
\be
{dE\over dt}={\dot R_H\over 2\,G}\,
{\sqrt{1-R_H/R+\dot R^2}-\dot R
\over\sqrt{1-R_H/R+\dot R^2}}
\simeq
{\dot R_H\over 2\,G}
\ ,
\label{dotRH}
\ee
one obtains, in the limit for $R$ close to $R_H$,
\be
\dot R_H\sim
-e^2\,G\,{N^2\over R^2}\,
\left({\ell_m\over R_H}\right)^3\,
\left(1-{R_H\over R}\right)^{1/2}\,\dot R^2
\label{RRH}
\ .
\ee
Were one to trust the above expression, one should then require that
the energy-momentum tensor of the radiation remain locally finite when
the shell approaches the surface $R=R_H^+$.
This amounts to the condition \cite{fulling},
\be
\dot R_H \sim \left(1-{R_H\over R}\right)^{\gamma}
\ ,
\ \ \ \ \
{\rm with}\ \gamma\ge 1
\ .
\label{reg1}
\ee
Since the quantity $B\sim 1$ near $R_H$, Eq.~(\ref{dotRH}) would lead
to the conclusion that the velocity of the macroshell satisfies
\be
\dot R^2\sim
\left(1-{R_H\over R}\right)^{\beta}
\ ,
\ \ \ \ \
{\rm with}\ \beta\ge {1\over 2}
\ ,
\ee
which means that the proper velocity $\dot R$ should vanish at the
horizon.
Of course, the number of approximations we have
employed does not allow one to consider the above a rigorous argument,
but just a suggestive result.
In particular, if the emission near the horizon is not thermal [$N_L$
in Eq.~(\ref{planck}) is small and just a few of the poles are included],
the flux satisfies the condition (\ref{reg1}) for any finite value
of $\dot R$, the backreaction is reduced and our results are significantly
modified only very close to the horizon where in any case our
approximations break down.
Indeed, we are only able to perform the analysis for $R$ greater than
a few times $R_H$ and, in all the cases we consider, the microshells
are confined with $\delta\ll R_H$.
\subsection{The trajectory}
\label{trajectory}
One can now integrate numerically the equation of motion for the radius
of the shell,
\be
\dot R^2=-1+{R_H^2\over(2\,G\,M)^2}+{R_H\over 2\,R}
+{G^2\,M^2\over 4\,R^2}
\ ,
\label{eqn1}
\ee
together with Eq.~(\ref{dotRH}) for $R_H$,
\be
\dot R_H\simeq
-{128\,\pi^3\over 9}\,\alpha\,
{G\,N^2\,\ell_m^{8/3}\over R_{\phantom H}^{10/3}\,R_H^{4/3}}\,
\left(1-{R_H\over R}\right)\,
{\dot R^2\over e^{2\,\Omega\,\sqrt{1-R_H/R}/k_B\,T_H}-1}
\ ,
\label{eqn2}
\ee
where $\alpha\equiv e^2/4\,\pi$.
\par
\begin{figure}
\centering
\raisebox{4cm}{$R$}\hspace{-0.2cm}
\includegraphics[width=2.9in]{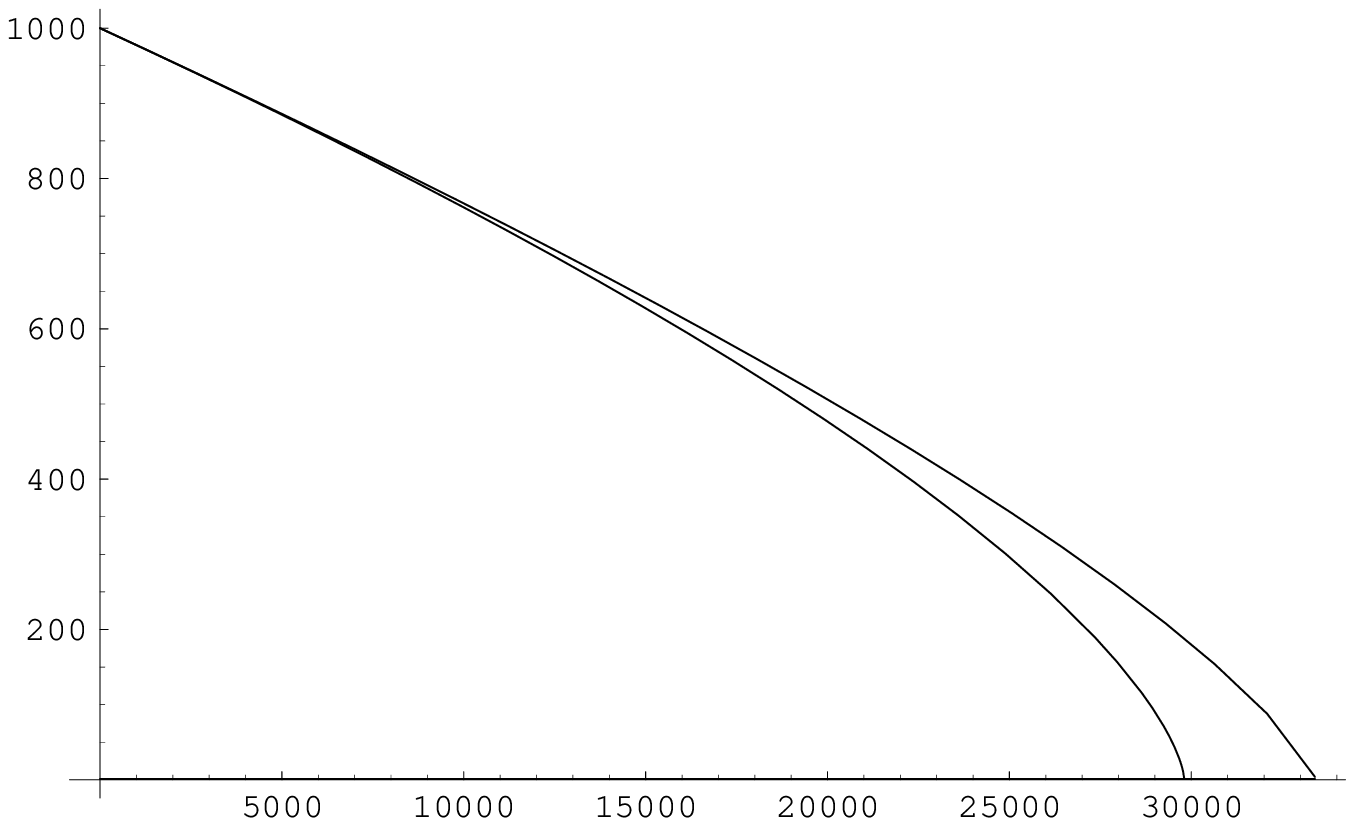}
\hspace{-0.2in}
\raisebox{4cm}{$R$}\hspace{-0.2cm}
\includegraphics[width=2.9in]{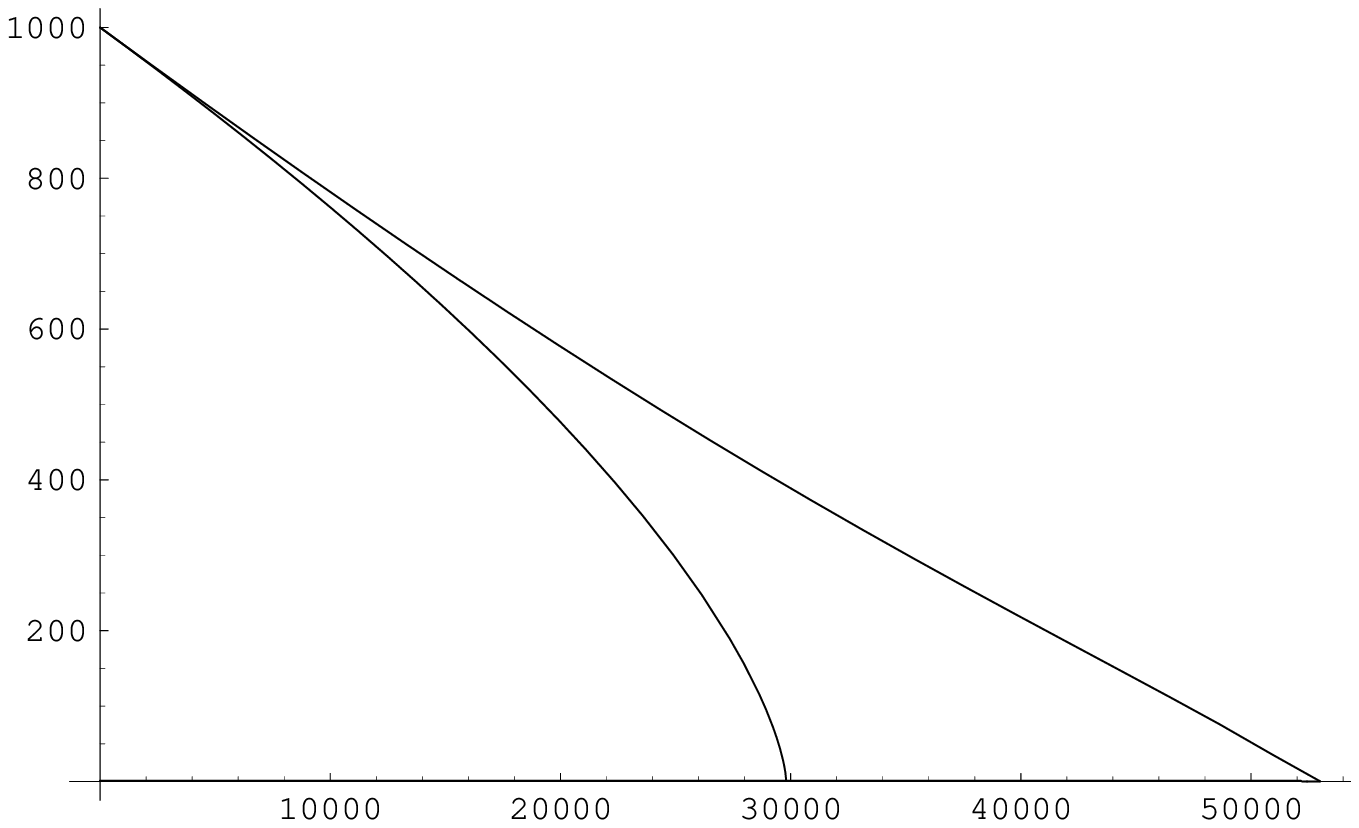}
\hspace{-0.2in}
\raisebox{0.5cm}{($\alpha=\alpha_0$)
\hspace{4cm}
$\tau$
\hspace{2cm}
($\alpha=4\,\alpha_0$)
\hspace{3cm}
$\tau$}\\
\raisebox{4cm}{$R$}\hspace{-0.2cm}
\includegraphics[width=2.9in]{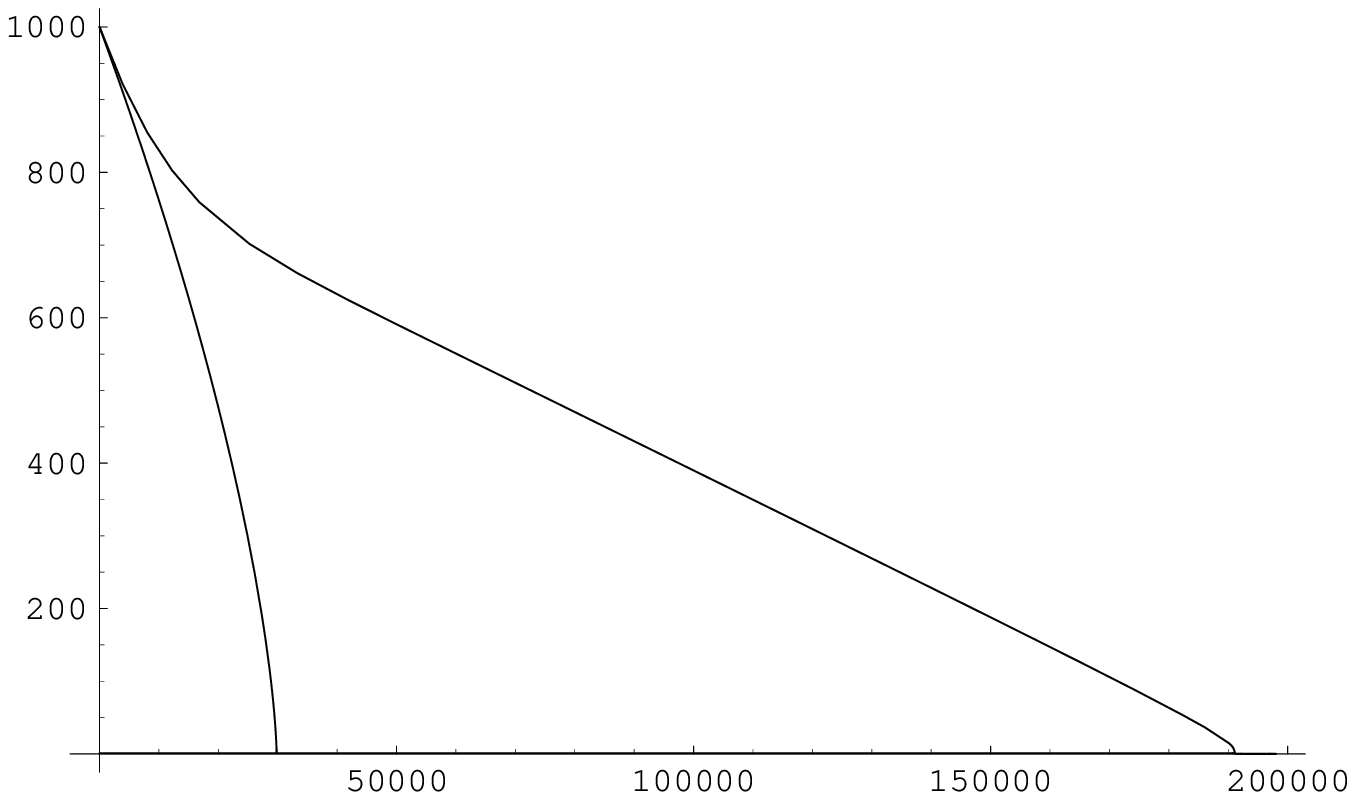}
\hspace{-0.2in}
\raisebox{4cm}{$R$}\hspace{-0.2cm}
\includegraphics[width=2.9in]{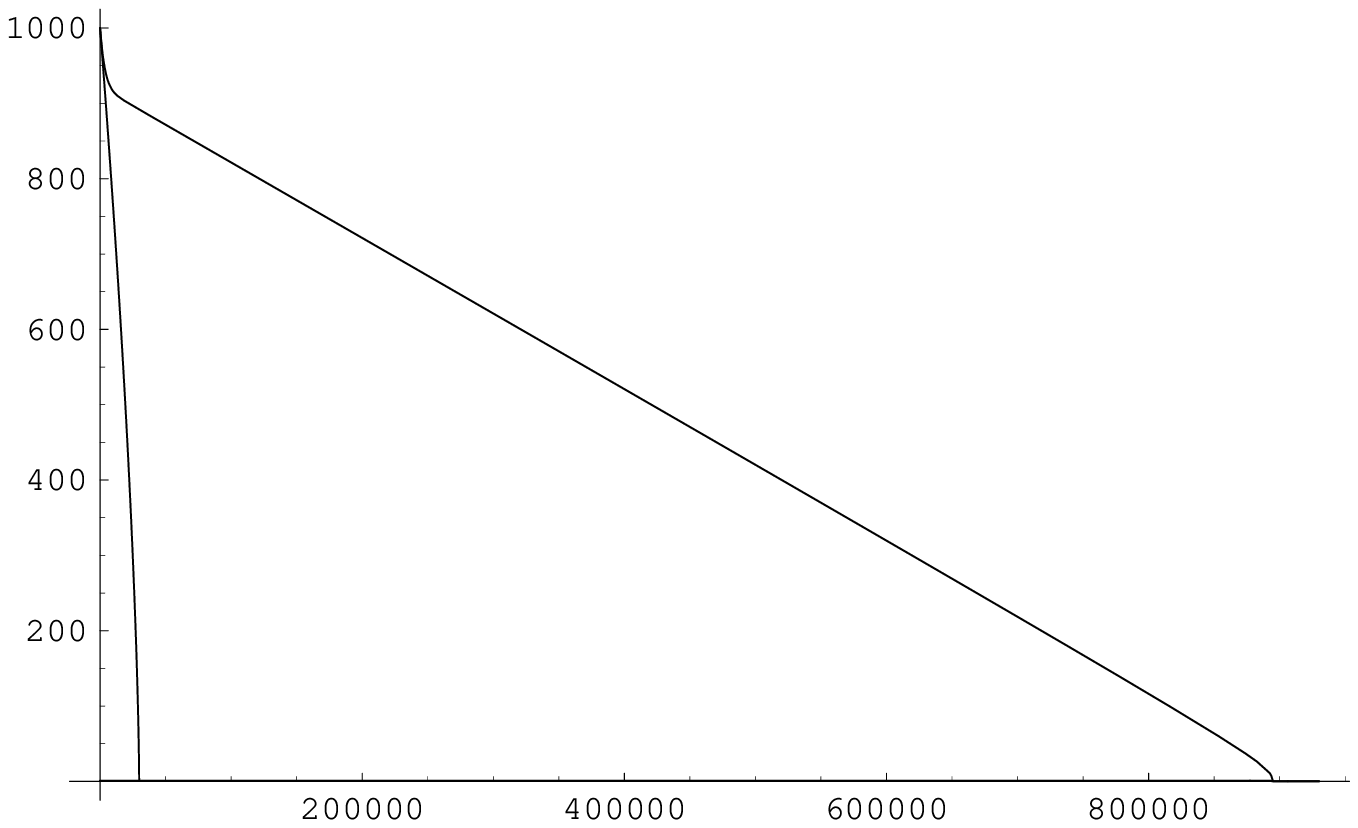}
\hspace{-0.2in}
\raisebox{0.5cm}{($\alpha=16\,\alpha_0$)
\hspace{4cm} $\tau$
\hspace{2cm}
($\alpha=64\,\alpha_0$)
\hspace{3cm} $\tau$}
\caption{Trajectories of the radiating shell $R=R(\tau)$ in units of
$R_H(0)$ with $N=2\cdot 10^{40}$ and four values of the coupling
constant $\alpha$ (upper curves) compared to the non-radiating collapse
(lower curves).
All trajectories are evolved from $R(0)=1000\,R_H(0)$.
The time $\tau$, in all plots, is expressed in
units of $2\cdot 10^{-38}\,N\,$GeV$^{-1}$ with $\hbar=c=1$ and
$\alpha_0=1.6\cdot 10^{-33}$.}
\label{RN2E40}
\end{figure}
\begin{figure}
\centering
{\hspace{6cm} $\tau$ \hspace{8cm} $\tau$}\\
\raisebox{4cm}{$R_H$}\hspace{-0.2cm}
\includegraphics[width=2.9in]{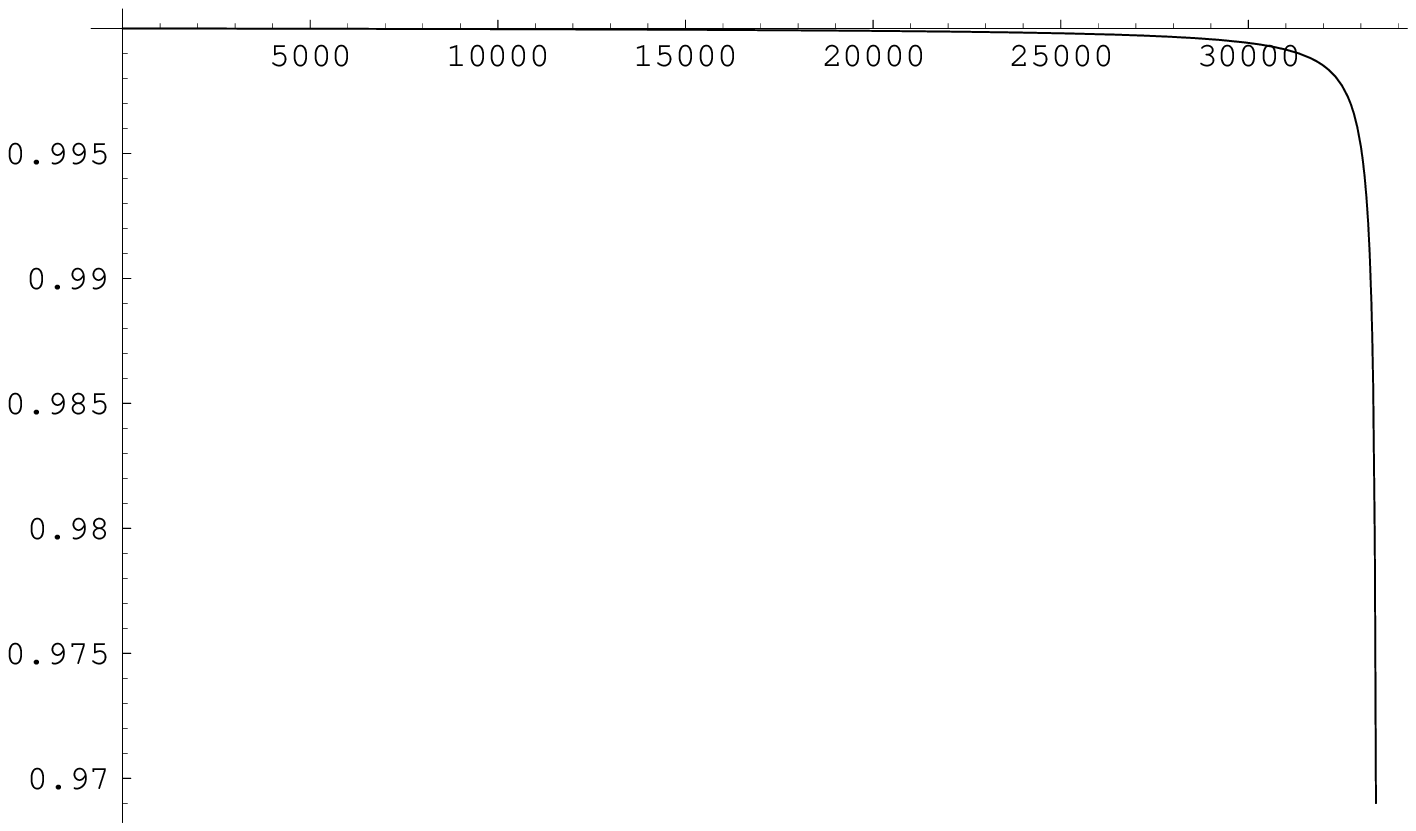}
\hspace{-0.0in}
\raisebox{4cm}{$R_H$}\hspace{-0.2cm}
\includegraphics[width=2.9in]{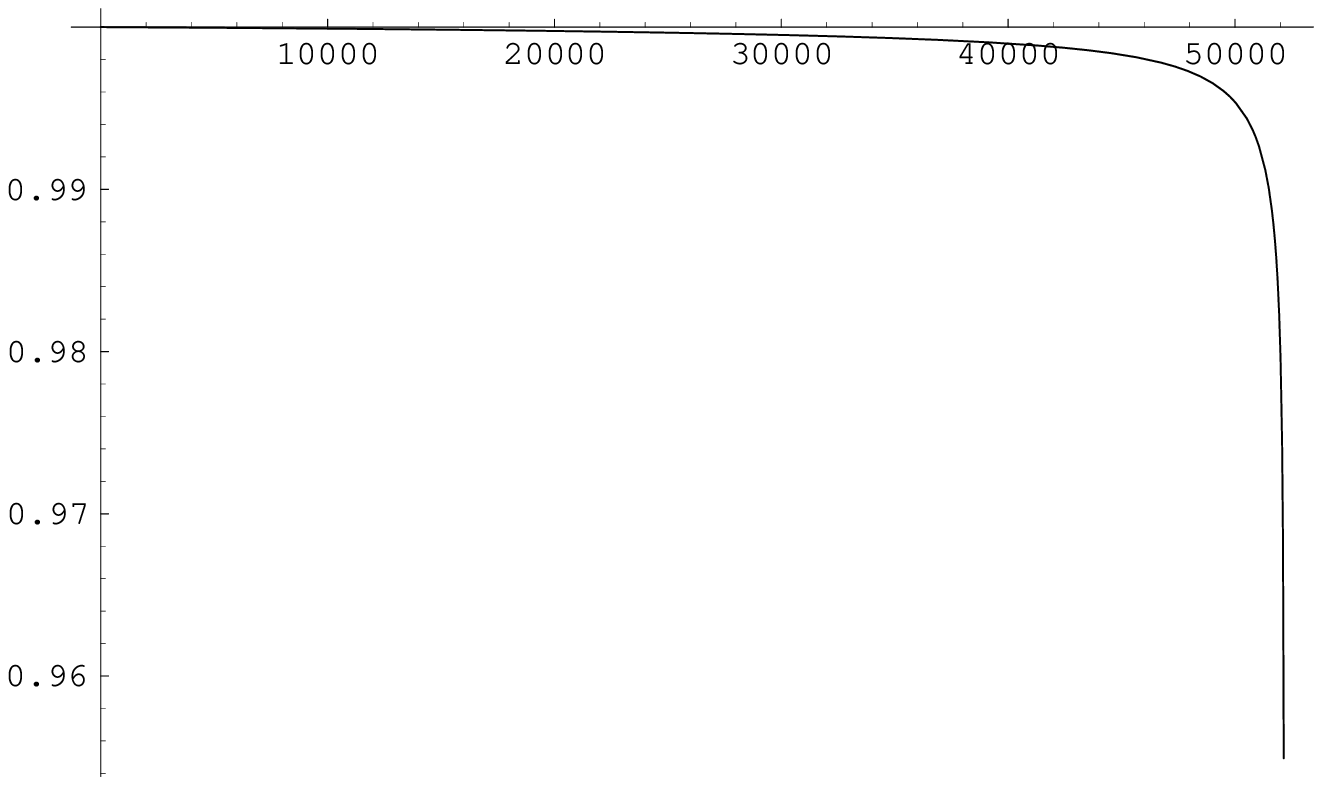}
\hspace{-0.2in}
\raisebox{0.5cm}{($\alpha=\alpha_0$)
\hspace{4cm}
\hspace{2cm}
($\alpha=4\,\alpha_0$)
\hspace{3cm}}\\
{\hspace{6cm} $\tau$ \hspace{8cm} $\tau$}\\
\raisebox{4cm}{$R_H$}\hspace{-0.2cm}
\includegraphics[width=2.9in]{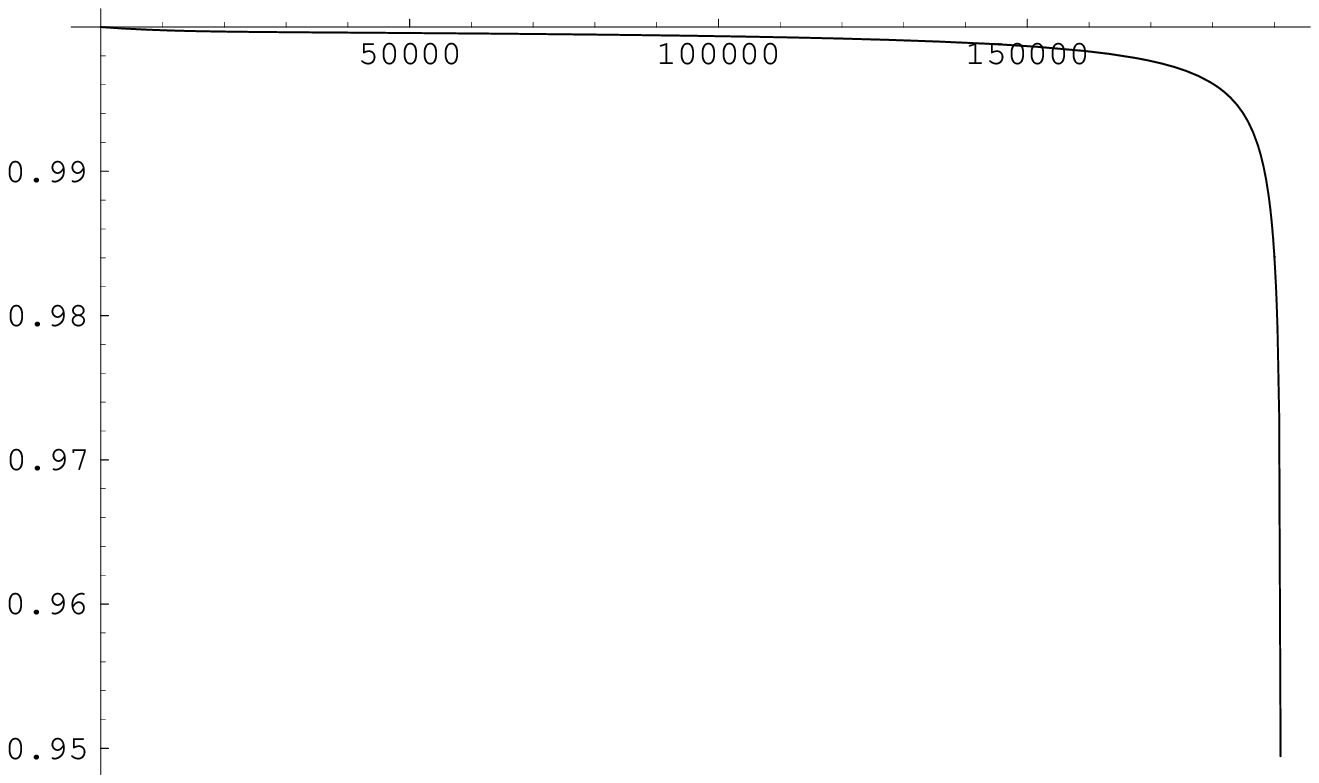}
\hspace{-0.0in}
\raisebox{4cm}{$R_H$}\hspace{-0.2cm}
\includegraphics[width=2.9in]{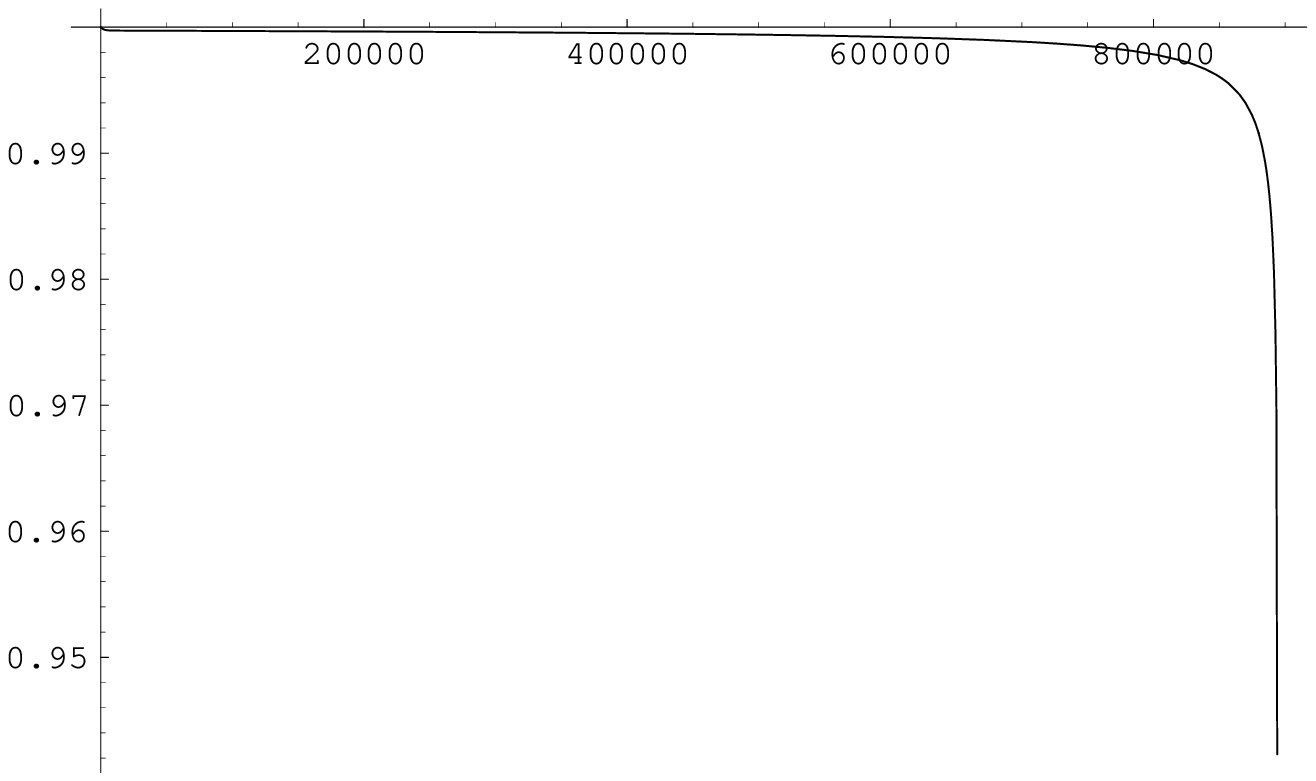}
\hspace{-0.2in}
\raisebox{0.5cm}{($\alpha=16\,\alpha_0$)
\hspace{4cm}
\hspace{2cm}
($\alpha=64\,\alpha_0$)
\hspace{3cm}}
\caption{Behaviour of the gravitational radius for the radiating
shell trajectories in Fig.~\ref{RN2E40}.}
\label{RhN2E40}
\end{figure}

Given our construction, the trajectories depend on the parameters $N$,
$\alpha$ and the initial condition $R(0)$.
In general, one finds that the non-adiabatic excitations occur relatively
close to the horizon and there is no strong dependence on the initial
value of $R$.
In Fig.~\ref{RN2E40} we display the comparison between the trajectories
computed from the above equations, with $N=2\cdot 10^{40}$ and various
values of $\alpha$, and that with constant $M_s=M$ (see also
Table~\ref{table1}).
Owing to our approximations, we are not able to display any reliable
results as $R$ gets close to $R_H$.
The behaviour of $R_H$ is shown in Fig.~\ref{RhN2E40}.
\begin{table}
\centerline{\begin{tabular}{|c|c|}
\hline
radiation wavelength $\lambda$ & $10^{-10}\div 10^{-7}\,$cm \\
\hline
$\delta/\lambda$ & $10^{-3}\div 10^{-1}$ \\
\hline
$\dot R$ & $10^{-3}\div 10^{-2}$ \\
\hline
total number of emissions & $10^{39}$ \\
\hline
$N_\delta$ & $10^{35}\div 10^{36}$ \\
\hline
total emitted energy & $5\%$ of $M_s(0)$
\\
\hline
\end{tabular}}
\caption{
Typical values of the relevant quantities resulting from the
numerical analysis for $N=2\cdot 10^{40}$ microshells
[$R_H(0)\simeq 4\cdot 10^{-12}\,$cm].
$N_\delta$ is the average number of emissions while the shell radius
shrinks a space $\delta$.}
\label{table1}
\end{table}
\par
\begin{figure}
\centering
\raisebox{4cm}{$R$}\hspace{-0.2cm}
\includegraphics[width=2.9in]{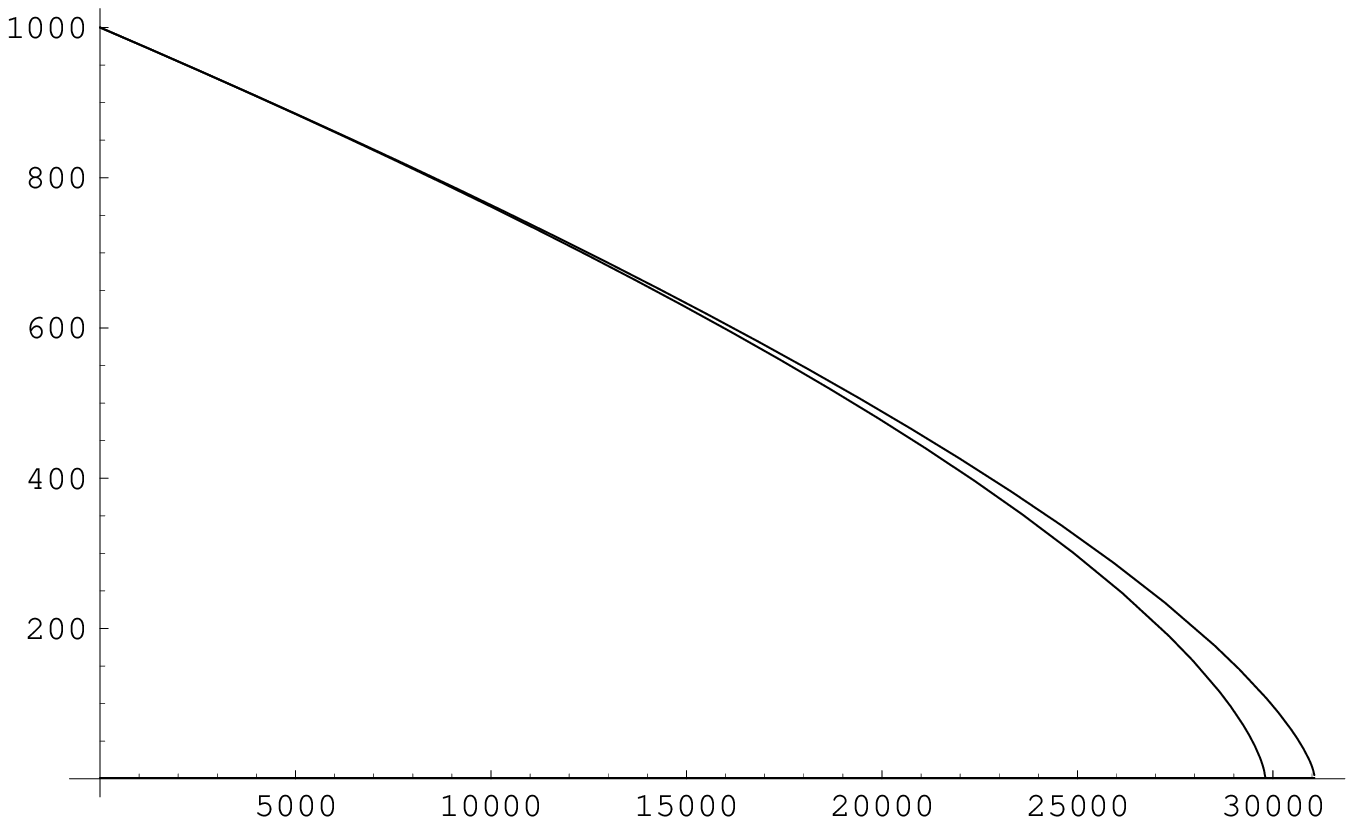}
\hspace{-0.2in}
\raisebox{4cm}{$R$}\hspace{-0.2cm}
\includegraphics[width=2.9in]{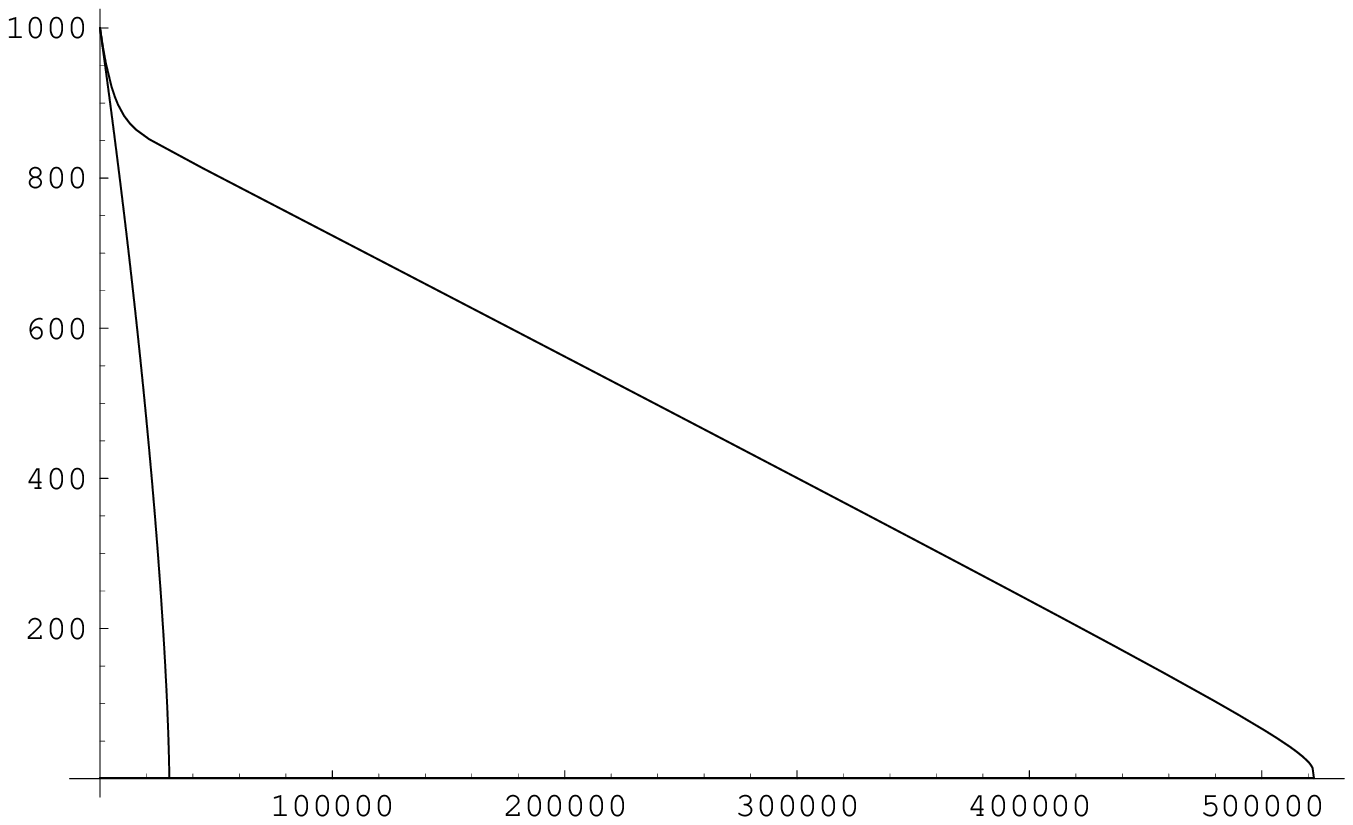}
\hspace{-0.2in}
\raisebox{0.5cm}{($\alpha=6.4\cdot 10^{-31}$)
\hspace{4cm} $\tau$
\hspace{2cm}
($\alpha=6.4\cdot 10^{-29}$)
\hspace{3cm}
$\tau$}
\caption{Trajectories of the radiating shell with $N=2\cdot 10^{41}$
and two values of the coupling constant $\alpha$ (upper curves)
compared to the non-radiating collapse (lower curves).}
\label{RN2E41}
\end{figure}

Increasing $N$ requires an increase in $\alpha$
in order to preserve the difference between the radiating and the
non-radiating trajectories (see Fig.~\ref{RN2E41} for a tenfold
increase in $N$ with respect to Fig.~\ref{RN2E40}).
In fact, from Eq.~(\ref{eqn2}), on setting $R\sim R_H\propto N\,m$
one finds (apart from dimensional constants)
\be
\dot R_H\propto{\alpha\over N^{8/3}\,m^{22/3}}
\ ,
\label{scale}
\ee
which gives an estimate of how the effect scales with respect to
the number of microshells, their mass $m$ and the radiation
coupling constant.
\par
\begin{figure}
\centering

{\hspace{6cm} $\tau$ \hspace{8cm} $\tau$}\\
\raisebox{4cm}{$\dot R$}\hspace{-0.2cm}
\includegraphics[width=2.9in]{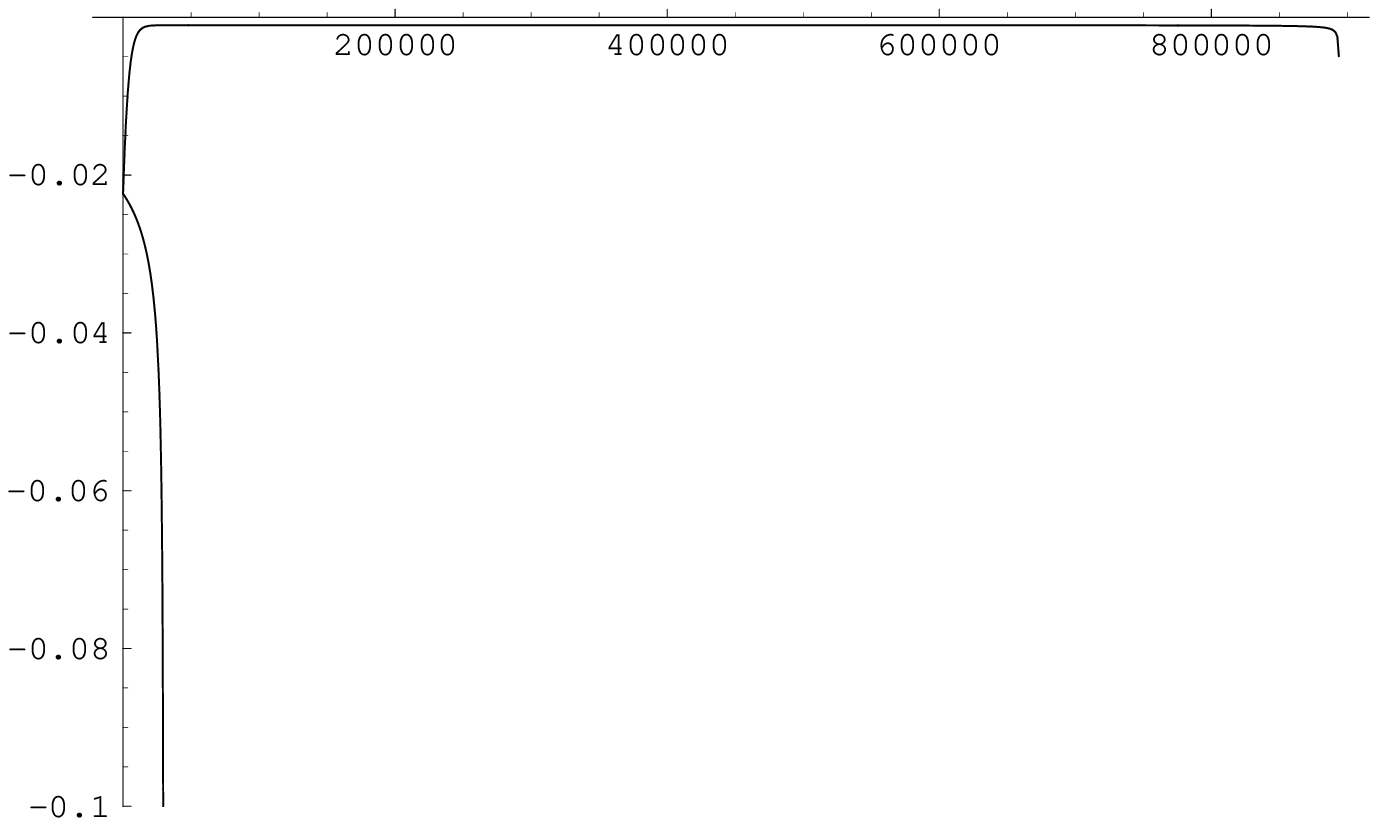}
\hspace{-0.0in}
\raisebox{4cm}{$\dot R$}\hspace{-0.2cm}
\includegraphics[width=2.9in]{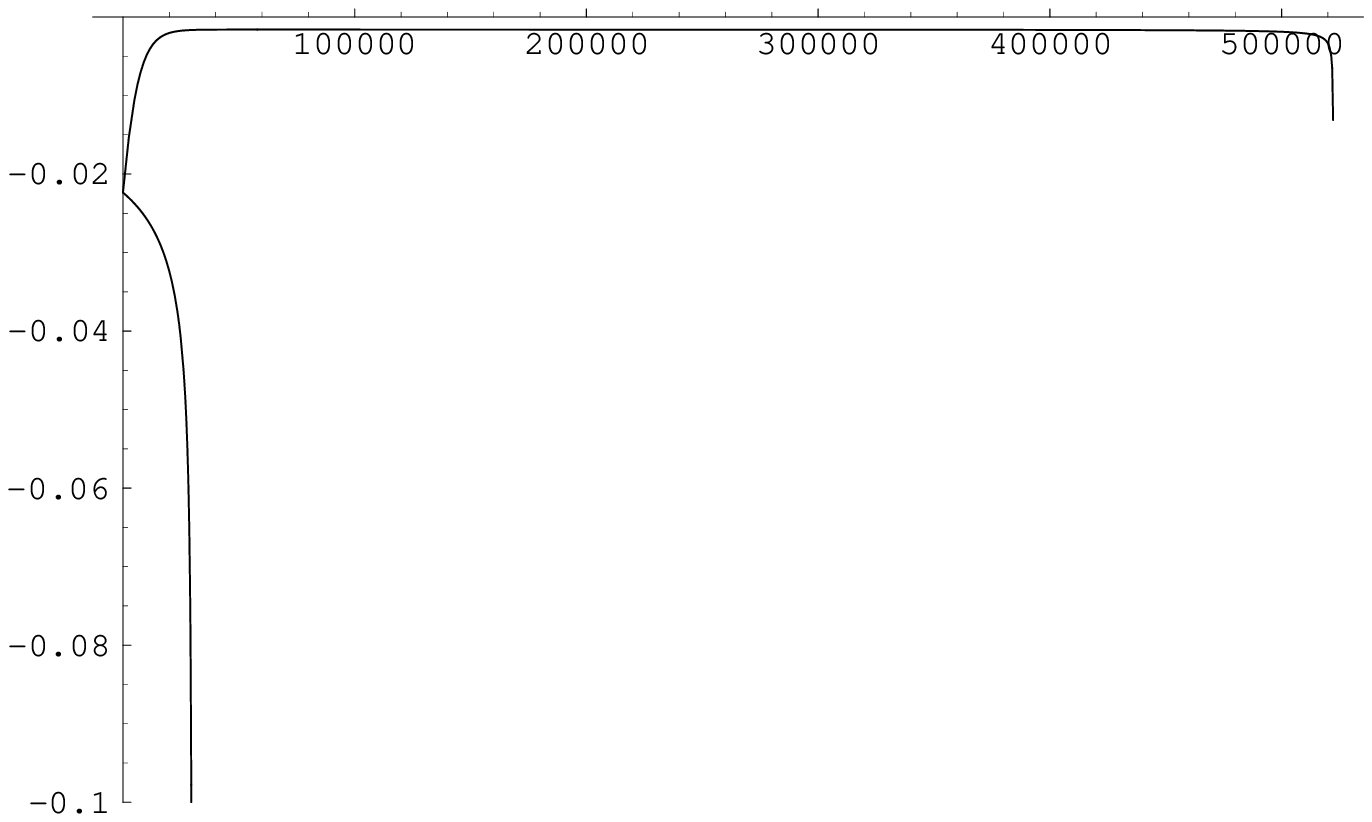}\\
{($N=2\cdot 10^{40}$, $\alpha=1\cdot 10^{-31}$)
\hspace{3cm}
($N=2\cdot 10^{41}$, $\alpha=6.4\cdot 10^{-29}$)}
\caption{Velocity of the radiating shell for the fourth case
of Fig.~\ref{RN2E40} and second case of Fig.~\ref{RN2E41}
(upper curves) compared to the non-radiating collapse
(lower curves).}
\label{vN2E41}
\end{figure}
\begin{figure}
\centering
\raisebox{4cm}{$\lambda$}\hspace{-0.2cm}
\includegraphics[width=2.9in]{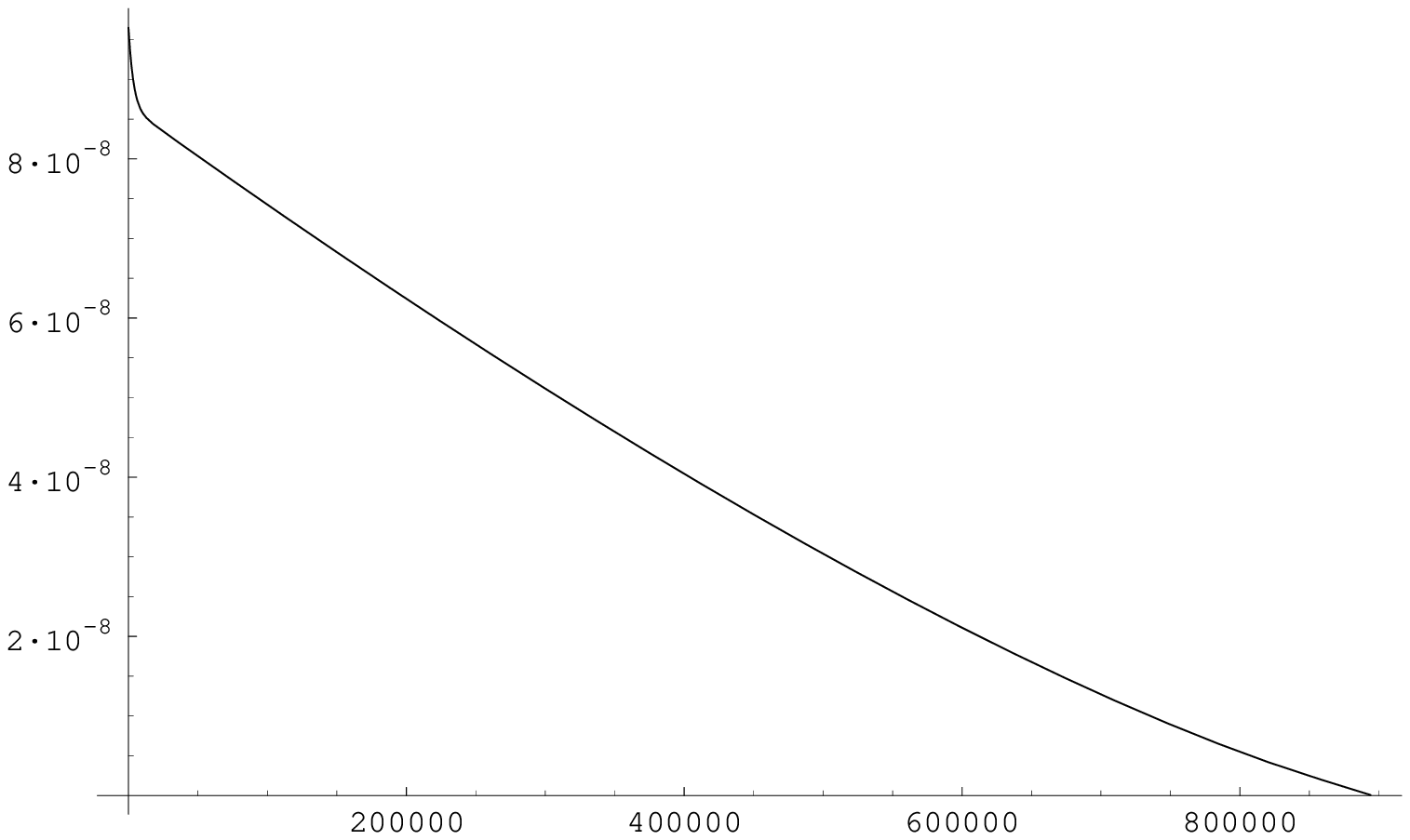}
\hspace{-0.0in}
\raisebox{4cm}{${\lambda\over \delta}$}\hspace{-0.2cm}
\includegraphics[width=2.9in]{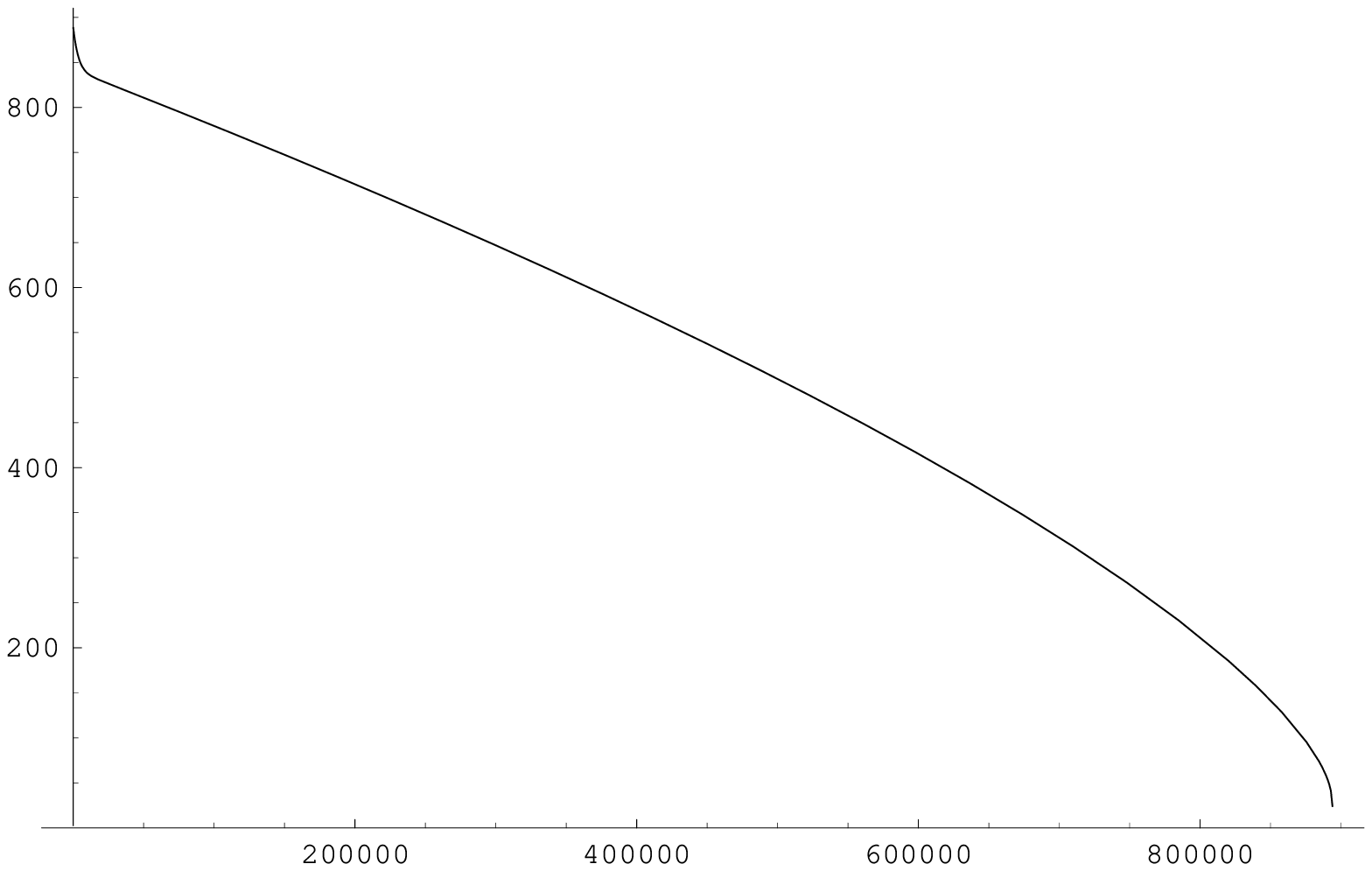}
\hspace{-0.2in}
\raisebox{0.5cm}{\hspace{6cm} $\tau$
\hspace{8cm}
$\tau$}
\caption{Wavelength $\lambda$ of the emitted quanta and
ratio $\lambda/\delta$ for the fourth case of Fig.~\ref{RN2E40}.}
\label{lN2E40}
\end{figure}

In Fig.~\ref{vN2E41} we plot the velocity $\dot R$ of the
collapsing shell (together with the velocity of the corresponding
non-radiating shell) for two relevant cases.
It is clear that $\dot R$ remains (negative and) small (within
a few percent of the speed of light), thus supporting our
approximation scheme within which we just retain the lowest order
in $\dot R$.
In Fig.~\ref{lN2E40} the wavelength of emitted radiation
quanta is plotted as a function of the time in order to show that it
remains larger than $\delta$ (as given in Table~\ref{table1}).
\par
Finally, we have also checked that on replacing the Bose-Einstein
factor in Eq.~(\ref{eqn2}) with the sum of just a few poles [see
Eq.~(\ref{planck})] the evolution of the system is not substantially
changed (only the backreaction is somewhat reduced).
\section{Conclusions}
\setcounter{equation}{0}
\label{conc}
In this paper we have analyzed the gravitational collapse
(in the semiclassical approximation) of a macroshell built up
of a (large) number of bosonic microshells ($s$-wave particles).
Starting from the classical equations of motion, we obtained
the potential and an effective Schr\"odinger equation for each
microshell.
The potential appearing in this equation ensures that the thickness
of the macroshell does not increase significantly for a long enough
piece of the trajectory, thus allowing one to take the thin shell
limit at some point in the calculations.
The time dependence of the potential leads to the non-adiabatic
excitations (which we computed to leading order in the velocity
of the macroshell) of the microshell bound states.
On coupling the microshells to a scalar radiation field, we showed
that the increase of kinetic energy can be radiated away and the
emitted radiation is approximately thermal with a temperature
given by the Tolman shifted instantaneous Hawking temperature
(whether this mechanism can be related to the Hawking effect
is a point which requires further study).
\par
The effect is intrinsically quantum mechanical, since it is a
consequence of the quantum mechanical--bound state nature of the
macroshell and the coherence of the emitted radiation.
This can cause the shell to lose enough energy so that the
backreaction on the trajectory of the radius is large.
For instance, for the case presented, the shell approaches
the gravitational radius in a time which is more than an order of
magnitude longer than the time a non-radiating shell would take
to cross the horizon, and loses about five percent of its ADM mass
as a burst of radiation (which is suggestive of the observed
gamma-ray bursts~\cite{tirado}).
One may also wonder at this point whether it is possible for the
collapsing matter to produce enough radiation to reduce the energy
of the shell so as to prevent the formation of an event horizon.
Thus, if the ADM mass decreased fast enough, the shell itself would
never reach the gravitational radius, unfortunately our approximations
become unreliable as it is approached.
\par
The model we have proposed clearly hinges on the formation
of a condensate.
Let us give some arguments in favour of this:
of relevance for the latter is the fraction of microshells
(particles) not in the condensate (ground state).
This fraction is related to $(T/T_c)^\nu$, where $T_c$ is the
critical temperature for Bose-Einstein condensation and
the (positive) exponent $\nu$ depends on the confining potential
and the dimensionality of the system~\footnote{For example,
$\nu=3/2$ for a three-dimensional box and $\nu=3$ for a
three-dimensional harmonic trap (for a review, see
Ref.~\cite{stringari}).}.
For our case of about $10^{40}$ particles and on setting $T=T_H$
(Hawking temperature) one has that $T_H/T_c<10^{-10}$
which renders our description in terms of a condensate
plausible.
\par
Naturally, the relevance of our bosonic model for gravitational
collapse which is expected to involve ordinary incoherent matter
may be questioned.
Moreover, it is clear that fermionic microshells would lead, because
of the Pauli exclusion principle, to a much wider macroshell with
and a loss of coherence.
However, for a mixed fermion-boson system, one may just consider the
bosonic part and apply our considerations to it since it will
condensate in the lowest state.
Certainly, the presence of fermionic and/or incoherent matter will
reduce the effects we have illustrated.
Nonetheless, one might even argue that an analogous phenomenon
can happen for all collapsing bosonic matter, including
the bosonic fraction of accretion disks around black holes.
Our result would thus suggest a new mechanism by which
the accreting matter can emit radiation.
\section*{Acknowledgement}
We thank R.~Brout for many enlighting discussions and
clarifications.
\appendix
\section{Microshell distribution and effective potential in NR}
\setcounter{equation}{0}
\label{newton}
The NR for each microshell of rest mass $m$ starting at large radius
($r_i\gg 2\,G\,M_s$) with negligible initial velocity can be obtained
by setting $M_>-M_<=m$ and keeping terms up to first order in
$G$ and to leading order in $m$ in the effective Hamiltonian constraint
(\ref{H_m}).
The total energy of the system is thus given by the sum of the kinetic
energies of all microshells and mutual interaction potential energies.
The latter can be written, on introducing relative ($\bar r_i$) and
centre-mass ($R$) coordinates, as \cite{shell}
\be
V(r_1,\ldots,r_N)=-G\,{M^2\over 2\,R}
+G\,{m^2\over 2\,R^2}\,\sum_{i<j}^{N}\,|\bar r_i-\bar r_j|
\ .
\ee
Unfortunately, this potential does not allow for an analytic treatment
of the complete problem, therefore we shall apply an analogous
procedure to that followed in Section~\ref{structure} in order to
obtain a Schr\"o\-din\-ger equation for each microshell in the mean
field of the others.
\par
Let us first try with a configuration $\{r_i,M_i\}_{hom}$ in which
$N-1$ microshells have equally spaced radii,
$r_i=r_1+(i-1)\,\delta/(N-1)$, so that
$M_i=m\,[(r_i-r_1)\,(N-1)/\delta+1]$.
Hence, the equation of motion (\ref{geo_x}) in NR for the $N^{th}$
microshell with radius $r_1<x<r_{N-1}$ becomes
\be
{1\over 2}\,m\,\left({dx\over dt}\right)^2
\simeq G\,{m^2\over x}\,\left(N\,{x-r_1\over \delta}+1\right)
\ ,
\ee
where $t\equiv \tau_x$ is the Newtonian time.
Changing to the relative coordinate $\bar x=x-R$, one obtains
\be
H_m\simeq {1\over 2}\,m\,\left({d\bar x\over dt}\right)^2
+G\,{m\,M\over R}\,{\bar x\over \delta}\,
\left({\bar x\over R}-1\right)
\ ,
\ee
where we have used $|\bar x|\ll R$ and omitted all potential terms
which do not contain $\bar x$ (they would just contribute to the
ground state energy in the Schr\"odinger equation).
It is clear that the potential above for $\bar x=\pm\delta/2$ does not
match smoothly with the NR limit of the external potential given in
Eq.~(\ref{linear_tau}) because of the term linear in $\bar x$.
This means that the configuration $\{r_i,M_i\}_{hom}$ is not
stable.
In fact, the minimum of the potential is not at the origin
$\bar x=0$, and one then expects the microshells to move away
from the above configuration.
\par
In order to obtain an internal configuration compatible with the
external potential, let us assume a potential with a minimum at
$\bar x=0$ (as is required by the symmetry of the external potential)
and consider small oscillations about this equilibrium point.
Such oscillations will be harmonic and the potential to lowest order
is given by
\be
V_{NR}=G\,{M\,m\over 2\,R^2\,\delta}\,(x-R)^2+C_1
\ ,
\ee
where $C_1$ is a constant.
If we assume $M_{i+1}=M_i+m$, for the most likely configuration
$\{r_i,M_i\}$, $r_i$ can be determined by comparing the potential
$V_{NR}$ with the one obtained from the NR limit of Eq.~(\ref{geo_x}),
to wit
\be
V_{NR}'=G\,{m\,(m-2\,M_i)\over 2\,x}+C_2
\ ,
\ \ \ \ \ r_i<x<r_{i+1}
\ ,
\ee
where $C_2$ is another constant.
On equating the potentials $V_{NR}$ and $V'_{NR}$ in the
limit $x\to r_i^+$ (from above) and setting $C=C_2-C_1$, one obtains
\be
2\,M_i-m=
r_i^+\,\left[{2\,C\over G\,m}
-{M\over R^2\,\delta}\,\left(r_i^+-R\right)^2\right]
\ .
\ee
Then, on defining $r_{i+1}=r_i+y_i$ and subtracting the same
expression for $x\to r_{i+1}^+$, one has
\be
2\,m&=&2\,M_{i+1}-2\,M_i
\nonumber \\
&=&{2\,C\over G\,m}\,y_i
+{M\,r_i^+\over R^2\,\delta}\,\left[\left(r_i^+-R\right)^2
-\left(r_i^++y_i-R\right)^2\right]
-{y_i\,M\over R^2\,\delta}\,\left(r_i^++y_i-R\right)^2
\nonumber \\
&=&{2\,C\over G\,m}\,y_i-{y_i\,M\over R^2\,\delta}\,\left[
R^2-4\,r_i^+\,R+3\,(r_i^+)^2+
y_i\,\left(3\,r_i^+-2\,R+y_i\right)
\right]
\ .
\label{step}
\ee
The first term on the right hand side corresponds to the homogeneous
distribution if the constant $C=G\,m^2\,(N-1)/\delta$.
In general, however, the other terms cannot be neglected, so that the
microshells are not equally spaced.
Further, the $y_i$ can be computed recursively for $i=1,2,\ldots,N-2$
on making use of Eq.~(\ref{step}) and the condition
$r_1\equiv R-\delta/2$ (which determines the constant $C$).
In any case, as we discuss in Appendix~\ref{schro}, the
actual quadratic or linear structure of the potential
inside the macroshell has little influence on the energy
levels of the lowest states.
\section{Shell dynamics in Schwarzschild time}
\setcounter{equation}{0}
\label{schwa}
It is known that, even if two theories are classically
equivalent, their quantization may lead to different
Hilbert spaces and, therefore, different quantum pictures
and that this problem is potentially present in Einstein's
gravity (see, {\em e.g.}, Ref.~\cite{hk}).
It is thus interesting to compare the potential $V_m^{(\tau)}$
obtained in Section~\ref{structure} with the one corresponding to a
foliation of space-time into slices parameterized by the
Schwarzschild time $t_\infty\equiv t_N$ measured by a (distant)
static observer.
Given this potential, one can (in principle) construct the Hilbert
space ${\mathcal{H}}_\tau$ and compute any observable in order to
check the equivalence of ${\mathcal{H}}_\tau$ and ${\mathcal{H}}_t$.
Unfortunately, the form of the potential we obtain in the following
makes the explicit construction practically impossible.
\par
First we observe that the proper time of the test shell can be
expressed in terms of the inner and outer Schwarzschild times
$t_<$ and $t_>\equiv t_x$ (see Fig.~\ref{macro}) by making use
of
\be
d\tau_x^2&=&\left(1-{2\,G\,M_<\over x}\right)\,dt_<^2
-\left(1-{2\,G\,M_<\over x}\right)^{-1}\,dx^2
\nonumber
\\
&=&\left(1-{2\,G\,M_>\over x}\right)\,dt_>^2
-\left(1-{2\,G\,M_>\over x}\right)^{-1}\,dx^2
\ ,
\label{dd}
\ee
which yields the ratio
\be
\left({dt_<\over dt_>}\right)^2=
{x-2\,G\,M_>\over x-2\,G\,M_<}
-{2\,x^2\,G\,(M_>-M_<)\over (x-2\,G\,M_>)\,(x-2\,G\,M_<)^2}\,
\left({dx\over dt_>}\right)^2
\ ,
\ee
where, from Eq.~(\ref{geo_x}),
\be
\left({dx\over dt_>}\right)^2=
{1\over x}\,{(x-2\,G\,M_>)^2\,F(x)\over
x-2\,G\,M_>+x\,F(x)}
\ .
\ee
These expressions can be iterated for each microshell with
radius $x<r_i\le r_N$ and, together with (\ref{dd}), yield
\be
\left({d\tau_x\over dt_\infty}\right)^2
&=&\left[\left(1-{2\,G\,M_x\over x}\right)
+\left(1-{2\,G\,M_x\over x}\right)^{-1}\,
\left({dx\over dt_x}\right)^2\right]
\nonumber \\
&&\times\prod_{i=x+1}^N\,\left[
{r_i-2\,G\,M_i\over r_i-2\,G\,M_{i-1}}
-{2\,r_i^2\,G\,(M_i-M_{i-1})\over
(r_i-2\,G\,M_i)\,(r_i-2\,G\,M_{i-1})^2}\,
\left({dr_i\over dt_i}\right)^2\right]
\nonumber \\
&=&\left[\left(1-{2\,G\,M_x\over x}\right)
+{(x-2\,G\,M_x)\,F(x)\over x-2\,G\,M_x+x\,F(x)}
\right]
\nonumber \\
&&\times\prod_{i=x+1}^N\,\left[
{r_i-2\,G\,M_i\over r_i-2\,G\,M_{i-1}}
+{2\,r_i^2\,G\,(M_i-M_{i-1})\,(r_i-2\,G\,M_i)\,F(r_i)\over
(r_i-2\,G\,M_{i-1})^2\,[r_i-2\,G\,M_i+r_i\,F(r_i)]}\right]
\ ,
\label{times}
\ee
where $r_{x+1}$ is the radius of the first microshell having radius
greater than that of the test shell and
\be
F(r_i)=-1+{(M_i-M_{i-1})^2\over m^2}
+G\,{M_i+M_{i-1}\over r_i}+G^2\,{m^2\over 4\,r_i^2}
\ .
\ee
On then multiplying Eq.~(\ref{geo_x}) by $m/2$ and using
Eq.~(\ref{times}), one obtains
the equation of motion for the microshell in the form of an
effective Hamiltonian constraint in the Schwarzschild time
\be
H_m^{(t)}\equiv {1\over 2}\,m\,\left({dx\over d\tau_x}\right)^2
\left({d\tau_x\over dt_\infty}\right)^2
+V^{(t)}=0
\ .
\ee
The explicit form of $V^{(t)}$ now depends both on the
distribution $\{r_i,M_i\}$ of the microshells inside the
macroshell and the cumbersome time conversion factor in
Eq.~(\ref{times}).
\par
Hence, we again rely on the thin shell limit ($\delta\ll r_1$)
and consider the form of the potential when $x<r_1$ or $x>r_N$.
For $x<X$ one finds [see case (1) in Fig.~\ref{ext}]
\be
\left\{\begin{array}{l}
\strut\displaystyle\left({dx\over dt_\infty}\right)^2=
\left[1-{2\,G\,m\over x}
+{(x-2\,G\,m)\,F_<(x)\over
x-2\,G\,m+x\,F_<(x)}\right]
\\
\phantom{\strut\displaystyle\left({dx\over dt_\infty}\right)^2=}\,
\times
\left[\strut\displaystyle{X-2\,G\,M_s\over X-2\,G\,m}
+{2\,X\,G\,(M_s-m)\,(X-2\,G\,M_s)\,F_<(X)\over
(X-2\,G\,m)^2\,[X-2\,G\,M_s+X\,F_<(X)]}
\right]\,F_<(x)
\\
\\
\strut\displaystyle\left({dX\over dt_\infty}\right)^2=
\left[1-{2\,G\,M_s\over X}
+{(X-2\,G\,M_s)\,F_<(X)\over
X-2\,G\,M_s+X\,F_<(X)}\right]
\,F_<(X)
\ .
\end{array}
\right.
\ee
Analogously, for $x>X$ [see case (2) in Fig.~\ref{ext}]
\be
\left\{\begin{array}{ll}
\strut\displaystyle\left({dX\over dt_\infty}\right)^2=
\left[1-{2\,G\,(M_s-m)\over X}
+{[X-2\,G\,(M_s-m)]\,F_>(X)\over
X-2\,G\,(M_s-m)+X\,F_>(X)}\right]
\\
\phantom{\strut\displaystyle\left({dX\over dt_\infty}\right)^2=}
\times\left[\strut\displaystyle
{x-2\,G\,M_s\over x-2\,G\,(M_s-m)}
+{2\,x\,G\,m\,(x-2\,G\,M_s)\,F_>(x)\over
[x-2\,G\,(M_s-m)]^2\,[x-2\,G\,M_s+x\,F_>(x)]}
\right]
\\
\phantom{\strut\displaystyle\left({dX\over dt_\infty}\right)^2=}
\times\,F_>(X)
\\
\\
\strut\displaystyle\left({dx\over dt_\infty}\right)^2=
\left[1-{2\,G\,M_s\over x}
+{(x-2\,G\,M_s)\,F_>(x)\over x-2\,G\,M_s+x\,F_>(x)}\right]
\,F_>(x)
\ ,
\end{array}
\right.
\ee
where $F_<$ and $F_>$ have been given in Eqs.~(\ref{F<}) and
(\ref{F>}).
On introducing the relative ($\bar r$) and centre-mass ($R$) radii,
after some lengthy algebra one obtains an effective Hamiltonian
for the two shells given by
\be
H^{(t)}={1\over 2}\,M\,\left({dR\over dt_\infty}\right)^2
+{1\over 2}\,\mu\,\left({d\bar r\over dt_\infty}\right)^2
+V^{(t)}\equiv H_M^{(t)}+H_m^{(t)}
\ ,
\ee
where the potential can be expanded in powers of $G$ and $\bar r/R$
according to
\be
V^{(t)}=\sum_{n=0}^\infty\,G^n\,\sum_{k=0}^\infty
V_{n,k}(M,m;M_s;R)\,\left({\bar r\over R}\right)^k
\ .
\ee
Only the terms linear in $\bar r/R$ are relevant for our
calculation and, to leading order in $G$, they are given by
\be
V_{1,1}=
\strut\displaystyle{M_s\,m\over 2\,R}\times
\left\{\begin{array}{ll}
\left(4\,\strut\displaystyle{M_s^2\over M^2}
-4\,{M_s\over M}-8+6\,{M\over M_s}
+6\,{M^2\over M_s^2}-2\,{M^3\over M_s^3}-{M^4\over M_s^4}
\right)
&
\ \ \bar r>+\strut\displaystyle{\delta\over 2}
\\
&
\\
\left(4\,\strut\displaystyle{M_s^2\over M^2}-10+
6\,{M^2\over M_s^2}-{M^4\over M_s^4}\right)
&
\ \ \bar r<-\strut\displaystyle{\delta\over 2}\ ,
\end{array}
\right.
\label{linear}
\ee
where we have again neglected non-leading terms in $m$.
\par
On interpolating between the two terms given by $V_{1,1}$
in Eq.~(\ref{linear}), we obtain
\be
H_m^{(t)}=
{1\over 2}\,m\,\left({d\bar r\over dt_\infty}\right)^2
+V_m^{(t)}
\ ,
\ee
with the potential
\be
V_m^{(t)}&=&G\,{M_s\,m\,\bar r^2\over 2\,R^2\,d}\,
\left(-2\,{M_s\over M}+1+3\,{M\over M_s}-{M^3\over M_s^3}\right)
\nonumber \\
&&+G\,{M_s\,m\,\bar r\over 2\,R^2}\,
\left(4\,{M_s^2\over M^2}-2\,{M_s\over M}-18
+3\,{M\over M_s}+6\,{M^2\over M_s^2}-{M^3\over M_s^3}
-{M^4\over M_s^4}\right)
\ ,
\label{c}
\ee
which holds for $|r|<\delta/2$.
\par
If we now take the thin shell limit of $V_m^{(t)}$ only the linear
part $V_{1,1}$ survives.
We can then compare the result with the corresponding linear
part of $V_m^{(\tau)}$, as computed in Section~\ref{structure},
using the Schwarzschild time, and then obtaining
\be
V_m^{(\tau)}(t)
=V_m^{(\tau)}\,\left({d\tau\over dt_\infty}\right)^2
\ ,
\ee
where
\be
\left({d\tau\over dt_\infty}\right)^2
=\left[1-{2\,G\,M_s\over R}+{(R-2\,G\,M_s)\,F(R)\over
R-2\,G\,M_s+R\,F(R)}\right]
\ ,
\ee
and
\be
F(R)=-1+{M_s^2\over M^2}+G\,{M_s\over R}+{G^2\,M^2\over 4\,R^2}
\ .
\ee
This yields, to lowest order in $G$,
\be
V_m^{(\tau)}(t)\simeq G\,{M_s\,m\,|\bar r|\over2\,R^2}\,
\left(2-{M^2\over M_s^2}\right)
\ ,
\ee
which coincides with the thin shell limit ({\em i.e.}, linear part)
of $V_m^{(t)}$, Eq.~(\ref{c}), for $M=M_s$.
This shows that the effective potential governing the motion of the
microshells is the same in NR for the two foliations $\{\Sigma_t\}$ and
$\{\Sigma_\tau\}$.
\begin{figure}
\centerline{\epsfxsize=300pt\epsfbox{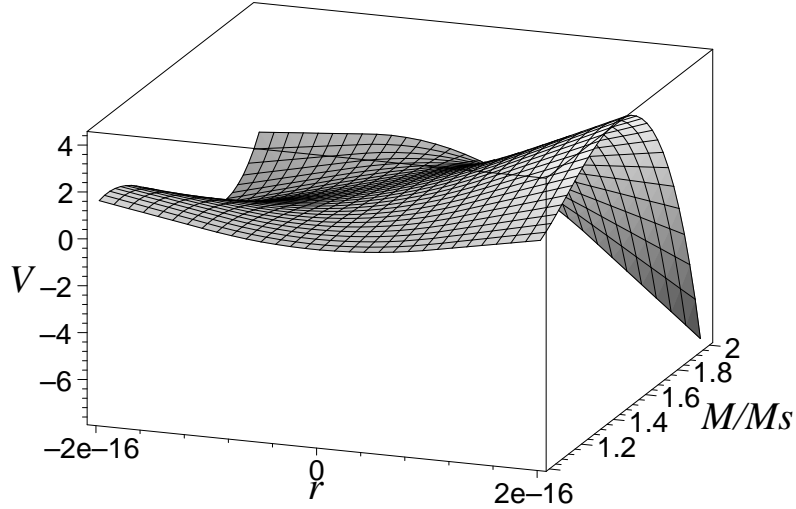}}
\caption{The potential $V_m^{(t)}$ at order $G^2$ for
$M=10^{40}\,m_h$ and $\delta=\ell_h=10^{-16}\,$m
$1\le M/M_s\le 2$.
Vertical units are arbitrary.}
\label{Vschw3D}
\end{figure}
\section{Microshell confinement}
\setcounter{equation}{0}
\label{schro}
\par
We shall now show that the expression for the macroshell thickness
$\delta$ in Eq.~(\ref{d}) can be obtained within the adiabatic
approximation ($R$ constant) in NR by either considering just the
linear part of the potential $V_m^{(\tau)}$ in $H_m^{(\tau)}$
(thus neglecting the details of the distribution of the microshells)
or just the quadratic part (which is related to a given distribution
of microshells, see Appendix~\ref{newton}).
From the Hamiltonian $H_m^{(\tau)}$ one can obtain a
(time-independent) Schr\"odinger equation
\be
\left[{\hat\pi_r^2\over 2\,m}\,
+V_{(i)}\right]\,\Phi_n=E_n\,\Phi_n
\ ,
\label{sc1}
\ee
where $\hat\pi_r=-i\,\hbar\,\partial_{\bar r}$ and, on retaining
the linear part, one has ($N\,m\equiv M=M_s$)
\be
V_{(1)}={G\,N\,m^2\over 2\,R}\,{|\bar r|\over R}
\ ,
\ee
or, on considering the quadratic term,
\be
V_{(2)}={G\,N\,m^2\over 2\, R^2}\,{\bar r^2\over\delta}
\ .
\ee
Solutions to Eq.~(\ref{sc1}) with $V_{(i)}=V_{(1)}$ are given
(apart from a normalization factor) by the Airy function
${\rm Ai}$ \cite{abra},
\be
\Phi_n^{(1)}=
\left\{\begin{array}{ll}
{\rm Ai}(\xi-\xi_{2\,k})
& \ \ \ \ \ n=2\,k
\\
& \\
{\rm sgn}(\bar r)\,{\rm Ai}(\xi-\xi_{2\,k+1})
& \ \ \ \ \ n=2\,k+1
\ ,
\end{array}\right.
\ee
where $k\in\natural$ and
$\xi\equiv(2\,G\,N\,m^3/\hbar^2\,R^2)^{1/3}\,\bar r$.
The $\xi_{2\,k+1}$ are the zeros of the Airy function,
${\rm Ai}(-\xi_{2\,k+1})=0$, and $\xi_{2\,k}$ the zeros of its
first derivative, ${\rm Ai}'(-\xi_{2\,k})=0$.
The corresponding eigenvalues are given by
\be
E_n^{(1)}=m\,\left({N\,\ell_p^2\over R^2}\right)^{2/3}
\,{\xi_n\over 2^{1/3}}
\ ,
\label{E1}
\ee
and the spread of such states (for $n$ small) is well approximated
by Eq.~(\ref{d}).
\par
\begin{figure}
\centerline{\epsfxsize=300pt\epsfbox{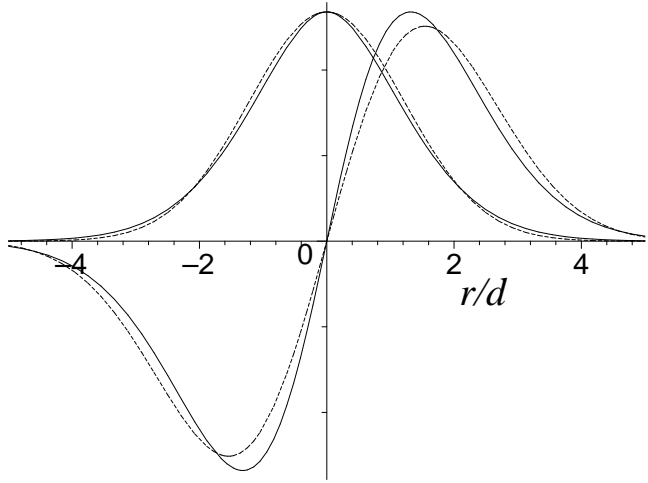}}
\caption{Plot of $\Phi_n^{(1)}$ (continuous lines) and $\Phi_n^{(2)}$
(dotted lines) for $n=0,1$.
Vertical units are arbitrary and $d$ should be $\delta$.}
\label{Airy}
\end{figure}

Solutions for $V_{(i)}=V_{(2)}$ are given instead by the usual
harmonic oscillator wavefunctions (again we omit normalization
factors),
\be
\Phi_n^{(2)}=H_n(\zeta)\,e^{-\zeta^2}
\ ,
\ee
where
$\zeta\equiv(G\,N\,m^3/2\,\hbar^2\,R^2\,\delta)^{1/4}\,\bar r$
and $H_n$ are Hermite polynomials \cite{abra}.
The energy eigenvalues are
\be
E_n^{(2)}=\sqrt{\hbar\,N\,m\over2\,\delta}\,{\ell_p\over R}
\,\left(n+{1\over 2}\right)
\ .
\ee
If we now substitute the expression for $\delta$ previously obtained,
we get the energy spectrum
\be
E_n^{(2)}=m\,\left({N\,\ell_p^2\over R^2}\right)^{2/3}\,
\left({n\over 2^{1/3}}+{1\over 2^{4/3}}\right)
\ ,
\ee
which is of the same order of magnitude as Eq.~(\ref{E1}).
Also, the spread of such $\Phi_n^{(2)}$ for $n$ small is again
given by Eq.~(\ref{d}) (see also Fig.~\ref{Airy}) and we can
therefore represent the state of each microshell either by using
$\Phi_n^{(1)}$ or $\Phi_n^{(2)}$.
\par
It is also important to observe that the energy gap between two
adjacent states,
\be
\Delta E\sim \left({N\,\ell_p^2\over R^2}\right)^{2/3}\,m
\sim \left({\ell_m\,R_H\over R^2}\right)^{2/3}\,m
\ ,
\ee
is much smaller than $m$ for $R\ge R_H$, which suggests that
outside the gravitational radius the effects considered in the
present paper dominate over the microshell creation studied in
Ref.~\cite{shell}.
\par
We may also see the small difference between the energy levels
in the two cases by employing a W.K.B. quantization \cite{shell}.
Indeed if we require
\be
\oint\pi_r\,d\bar r=m\,\oint \dot{\bar r}\, d\bar r
=\left(n+{1\over 2}\right)\,h
\ ,
\ee
for both potentials and consider cycles within the potential well,
that is the amplitude of the oscillations will correspond to the
width of the shell, one has
\be
m\,\oint \dot{\bar r}\, d\bar r=
\sqrt{2\,m}\,\int_{-\delta/2}^{+\delta/2} d\bar r\,
\left[V_{(i)}(\delta/2)-V_{(i)}(\bar r)\right]^{1/2}
=\left(n+{1\over 2}\right)\,h
\ ,
\ee
thus determining $\delta_{(1)}$ and $\delta_{(2)}$ for the linear
and quadratic potentials respectively.
One obtains
\be
\delta_{(1)}&=&
2\,\left[{3\over 2}\,\left(n+{1\over 2}\right)\,h\,
\left({R^2\over 16\,G\,M\,m}\right)^{1/2}\right]^{2/3}
\nonumber \\
&&\\
\delta_{(2)}&=&
2\,\left[{4\,\sqrt{2}\over \pi}\,\left(n+{1\over 2}\right)\,h\,
\left({R^2\over 16\,G\,M\,m}\right)^{1/2}\right]^{2/3}
\ ,
\nonumber
\ee
and correspondingly a binding energy
\be
E^{(1)}_n&=&\left({9\over 128}\right)^{1/3}\,
\left[\left(n+{1\over 2}\right)\,h\right]^{2/3}\,
\left({G\,M\over 2\,R^2}\right)^{2/3}\,m
\nonumber \\
&& \\
E_n^{(2)}&=&\left({1\over \pi^2}\right)^{1/3}\,
\left[\left(n+{1\over 2}\right)\,h\right]^{2/3}\,
\left({G\,M\over 2\,R^2}\right)^{2/3}\,m
\ ,
\nonumber
\ee
which are of the same order of magnitude and we have added to
$E_n^{(2)}$ the additional ${\cal O}(G)$ term in
Eq.~(\ref{inner_tau}) so as to have continuity of the potentials
at $|\bar r|=\delta/2$.
It is clear that in both cases we have allowed for a maximal
oscillation of the test microshell and thus obtained $\delta_{(i)}$.
\par
Finally, let us now check how close to the gravitational radius
we can use the bound states obtained above.
From Eq.~(\ref{linear_tau}) one finds that $V^{(\tau)}$ is always
positive for $\bar r<0$, but becomes negative for $\bar r>0$
if
\be
R<{G^2\,M^2\over 2\,R_H}\equiv R_C
\ ,
\label{deco}
\ee
As a sufficient condition for applicability of the modes
$\Phi_n^{(i)}$, we shall require that the energy be less than
the value of the potential for $\bar r\sim 4\,\delta$
(see Fig.~\ref{Airy}), {\em i.e.},
$E_n^{(i)}< V_{(1)}(\bar r=4\,\delta)$.
The above condition implies $n<4$ for $R>R_H$ ($>R_C$)
and is therefore satisfied for the second excited state ($n=2$)
we use in Section~\ref{f&t}, otherwise the shell breaks up.
\end{document}